\documentclass[useAMS]{mn2e}
\usepackage{graphicx}
\usepackage{txfonts}
\newcommand\aj{AJ}
\newcommand\apj{ApJ}
\newcommand\apjs{ApJS}

\newcommand\aap{A\&A}
\newcommand\mnras{MNRAS}
\newcommand\apjl{ApJ}
\newcommand\pasp{PASP}

\newcommand\nat{Nature}

\title[]{The {\it Hubble Space Telescope} UV Legacy Survey of Galactic Globular Clusters. IX. The Atlas of Multiple Stellar Populations.}
\author[Milone et al.]
       {A.\,P.\,Milone$^{1}$,
         G.\,Piotto$^{2,3}$,
         A.\,Renzini$^{3}$,
         A.\,F.\,Marino$^{1}$,
         L.\,R.\,Bedin$^{3}$, 
         E.\,Vesperini$^{4}$, \newauthor
         F.\,D'Antona$^{5}$,
         D.\,Nardiello$^{2}$,
         J.\,Anderson$^{6}$,
         I.\,R.\,King$^{7}$,
         D.\,Yong$^{1}$,
         A.\,Bellini$^{6}$,\newauthor
         A.\,Aparicio$^{8, 9}$,
         B.\,Barbuy$^{10}$,         
         T.\,M.\,Brown$^{6}$, 
         S.\,Cassisi$^{11}$,
         S.\,Ortolani$^{2}$,\newauthor 
         M.\,Salaris$^{12}$,
         A.\,Sarajedini$^{13}$,
         R.\,P.\,van der Marel$^{6}$\\
$^{1}$Research School of Astronomy \& Astrophysics, Australian National University, Mt Stromlo Observatory, via Cotter Rd, Weston, ACT 2611, Australia \\
$^{2}$Dipartimento di Fisica e Astronomia ``Galileo Galilei'', Univ. di Padova, Vicolo dell'Osservatorio 3, Padova, IT-35122\\
$^{3}$Istituto Nazionale di Astrofisica - Osservatorio Astronomico di Padova, Vicolo dell'Osservatorio 5, Padova, IT-35122\\
         $^{4}$Department of Astronomy, Indiana University, Bloomington, IN 47405, USA\\
         $^{5}$Istituto Nazionale di Astrofisica - Osservatorio Astronomico di Roma, Via Frascati 33, I-00040 Monteporzio Catone, Roma, Italy\\
         $^{6}$Space Telescope Science Institute, 3800 San Martin Drive, Baltimore,  MD 21218, USA\\
$^{7}$Department of Astronomy, University of Washington, Box 351580, Seattle, WA 98195-1580\\
         $^{8}$Instituto de Astrof\`isica de Canarias, E-38200 La Laguna, Tenerife, Canary Islands, Spain\\
$^{9}$Department of Astrophysics, University of La Laguna, E-38200 La Laguna, Tenerife, Canary Islands, Spain\\
         $^{10}$ Universidade de Sao Paulo, IAG, Rua de Matao 1226, Ciudade Universitaria. Sao Paulo 05508-900, Brazil\\
$^{11}$Istituto Nazionale di Astrofisica - Osservatorio Astronomico di Teramo, Via Mentore  Maggini s.n.c., I-64100 Teramo, Italy\\
$^{12}$Astrophysics Research Institute, Liverpool John Moores University, Liverpool Science Park, IC2 Building, 146 Brownlow Hill, Liverpool L3 5RF, UK\\
$^{13}$Department of Astronomy, University of Florida, 211 Bryant Space Science Center, Gainesville, FL 32611, USA\\
       }

\begin{document}
%\date{Accepted xxx December 15. Received xxx December 14; in original form xx October 11}
\date{Draft Version September, 30, 2016}

\pagerange{\pageref{firstpage}--\pageref{lastpage}} \pubyear{2016}

\maketitle
\label{firstpage}
 
\begin{abstract}  
We use high-precision photometry of red-giant-branch (RGB) stars in 57 Galactic globular clusters (GCs), mostly from  the ``{\it Hubble Space Telescope} ({\it HST\,}) UV Legacy Survey of Galactic globular clusters",  to identify and characterize their multiple stellar  populations.
 For each cluster the pseudo two-color  diagram (or `chromosome map')  is presented, built with a suitable  combination of stellar magnitudes in the F275W, F336W, F438W and F814W filters that maximizes the separation between multiple populations. 
 In the chromosome map of most GCs (Type I clusters), stars separate in two distinct groups that we identify with the first (1G) and the second generation  (2G). This identification is further supported by noticing that 1G stars have primordial (oxygen-rich, sodium-poor) chemical composition, whereas 2G stars are enhanced in sodium  and depleted in oxygen. This 1G-2G separation is not possible for a few GCs  where the two sequences have apparently merged into an extended, continuous sequence.
In some GCs (Type II clusters) the 1G and/or the 2G sequences appear to be split, hence displaying more complex chromosome maps. These clusters exhibit 
 multiple SGBs also in purely optical color-magnitude diagrams, with the fainter SGB joining into a red RGB which   is populated by stars with enhanced heavy-element abundance. We measure the RGB width by using appropriate colors and pseudo-colors.  
 When the metallicity dependence is removed, the RGB width correlates with the cluster mass. 
The fraction of 1G stars ranges from $\sim$8\% to $\sim$67\% and anticorrelates with the cluster mass,  indicating that incidence and complexity of the multiple population phenomenon both increase with cluster mass.

\end{abstract}


\section{Introduction}\label{sec:intro}
The formation of globular clusters (GCs) and the origin of their ubiquitous  multiple stellar populations remain a major astrophysical challenge. In this series of papers, we build on the {\it Hubble Space Telescope} ({\it HST\,}) UV Legacy Survey of Galactic Globular Clusters (Piotto et al. 2015, hereafter Paper I of this series) to fully document the complexity of the multiple populations. This  phenomenon is most effectively characterized when combining ultraviolet and optical  {\it HST} photometry, as documented by pilot studies by our group 
(e.g., Milone et al.\,2012b, 2013; Piotto et al.\,2013). These studies have demonstrated that appropriate combinations of ultraviolet and optical filters, to construct e.g., $m_{\rm F275W}-m_{\rm F336W}$ vs.\,$m_{\rm F336W}-m_{\rm F438W}$ two-color diagrams or the $m_{\rm F814W}$ plot vs. the pseudo-color $C_{\rm F275W,F336W,F438W} = (m_{\rm F275W}-m_{\rm F336W})-(m_{\rm F336W}-m_{\rm F438W}$), very efficiently identify multiple stellar populations in GCs  (see Paper\,I for a general introduction into the subject).

In other papers (Milone et al.\,2015a,b hereafter Papers II and III), we have shown that the %$\Delta_{\rm F275W,F336W,F438W}$ vs.\,$\Delta_{\rm F275W,F814W}$ 
 combination of the $C_{\rm F275W,F336W,F438W}$ pseudo-color with the $m_{\rm F275W}-m_{\rm F814W}$ color maximizes the separation between  stellar populations along the main sequence (MS) and the red giant branch (RGB) and have used this diagram to identify and characterize seven distinct stellar populations in NGC\,7089 (Paper II) and at least five populations in NGC\,2808 (Paper III). In Paper IV of this series we provided accurate determination of the GC helium abundance and ages of stellar populations in NGC\,6352 (Nardiello et al.\,2015a), while in Paper V we have exploited the first results from our survey to set constraints on the formation scenarios (Renzini et al.\,2015). 
 Other papers of this series include the study of the internal dynamics of multiple populations (Bellini et al.\,2015, Paper VI) and of the horizontal branch (HB) morphology (Brown et al.\,2016, Paper VII). An early-stage data release of the photometric and astrometric data  is provided in Paper VIII by Soto et al.\,(2016).

 In this paper, we identify and characterize multiple stellar populations along the RGB for the entire sample of 57 GCs. The paper is organized as follows.
 Data reduction and analysis are briefly described in Section~\ref{sec:data}.
In Section~\ref{sec:map} we measure the intrinsic RGB width in $C_{\rm F275W,F336W,F438W}$ and $m_{\rm F275W}-m_{\rm F814W}$  for all the clusters and we describe how to combine these two quantities to construct  `chromosome maps', which most efficiently  identify the distinct stellar populations  hosted by each individual GC. The chromosome maps of all the clusters are presented  in Section~\ref{sec:rea}. We distinguish between putative first and second generations of stars (respectively 1G and 2G) and measure the fraction of 1G stars over the total cluster population.  A group of GCs exhibiting particularly complex chromosome maps and characterized by the presence of a multimodal subgiant branch (SGB) are further investigated in  Section~\ref{sec:ano}. In Section~\ref{sec:relW}  we  present univariate relations between the  global parameters of the host clusters, the RGB  width and the population ratio.  
Summary and conclusions follow in Section~\ref{sec:summary}.

\section{Data and data analysis}
\label{sec:data}

This study is mostly based on data from the {\it HST} program GO-13297  (PI.\,G.\,Piotto) and data from the pilot programs GO-12605 and GO-11233 from the same PI. The aim of these programs is to derive high-precision photometry and astrometry of stars in 57 clusters through the F275W, F336W, and F438W filters of the  of  {\it HST} Ultraviolet and Visual Channel of the Wide Field Camera 3 (WFC3/UVIS).  In addition to data and catalogs illustrated in  Paper\,I , we make use of F606W and F814W photometry from the Wide Field Channel of the Advanced Camera for Survey (WFC/ACS) which is available for all clusters, mainly from the ACS survey of Galactic Globular Clusters (GO-10775, PI.\,A.\,Sarajedini, see Sarajedini et al.\,2007).
In order to improve the quality of the photometry for a few clusters, we have included additional archival WFC3/UVIS images in F275W, F336W, and F438W  as reported  in Table~\ref{tab:data}.

 All the images have been corrected for the effect of poor charge transfer efficiency following Anderson \& Bedin\,(2010).  Photometry has been performed on each individual exposure by using the program img2xym\_wfc3uv, which has been developed by Jay Anderson and is similar to the img2xym\_WFC program  (Anderson \& King\,2006), but optimized for UVIS/WFC3 data. 
For saturated very-bright stars the  photometry  was performed using  the method developed by Gilliland (2004), which recovers the electrons that have bled into neighbouring pixels. We refer to Section~8.1 in Anderson et al.\,(2008) for details on the application of this method.

 Stellar positions have been corrected for geometric distortion using the solution by Bellini, Anderson \& Bedin\,(2011). Photometry has been calibrated to the Vega-mag system as in Bedin et al.\,(2005), by using the photometric zero points provided by the WFC3/UVIS web page\footnote{http://www.stsci.edu/hst/acs/analysis/zeropoints/zpt.py}.  Stellar proper motions have been obtained as in Anderson \& King (2003) and Piotto et al.\,(2012) by comparing the average stellar positions derived from the WFC3 images in the F336W and F438W bands with those from the catalogs by Anderson et al.\,(2008). We have included in the following analysis only stars that are cluster members according to their proper motions. 

 Since we are interested in high-precision photometry, we limited our study to relatively-isolated stars with small astrometric uncertainties that are well fitted by the PSF and  selected by following the prescriptions given in Milone et al.\,(2012b).  Finally, the photometry has been corrected for differential reddening  that area crucial step in identifying multiple sequences from photometry. To do this we have used the method by Milone et al.\,(2012b). In a nutshell, we derived a $\sigma$ clipped fiducial line of the MS and the SGB of each cluster by putting a spline through the median value of colors and magnitude in progressively narrower magnitude intervals.  We have then determined for each star the color residuals of the closest 50 relatively-bright and well-measured MS and SGB stars with respect to the fiducial line. To do this we have excluded the target star from the calculation of its own differential reddening. We assumed as the differential reddening of each star the median value of such residuals measured along the reddening vector, while the uncertainty on the differential-reddening  has been derived as in Milone\,(2015). 

\begin{table*}
\caption{Description of the archive HST images set that have been used in this paper in addition to GO-10775, GO-11233, GO-12605, and GO-13297 data.}
\begin{tabular}{ccccccc}                                                                                    
\hline\hline
CLUSTER & DATE & N$\times$EXPTIME & FILTER & INSTRUMENT & PROGRAM & PI \\
\hline
 NGC\,0104 & Sep 30 -- Nov 10 2002 & 3$\times$150s$+$6$\times$100s$+$10s  & F435W & WFC/ACS & 9281 & J.\,Grindlay \\
 NGC\,0104 & Sep 30 -- Jul 07 2002 & 100s  & F435W & WFC/ACS & 9443 & I.\,King \\
 NGC\,0104 & Sep 28 -- 29 2010 & 2$\times$580s$+$30s  & F336W   & WFC3/UVIS & 11729 &  W.\,Freedman\\
 NGC\,0104 & Nov 14 2012 -- Sep 20 2013 & 9$\times$485s$+$9$\times$720s  & F336W & WFC3/UVIS & 12971 & H.\,Richer \\
 NGC\,5139 & Jul 15 2009 & 9$\times$350s+35s & F275W & WFC3/UVIS & 11452 & J.\,K.\,Quijano\\
 NGC\,5139 & Jan 12 -- Jul 04 2010 & 22$\times$800s & F275W & WFC3/UVIS & 11911 & E.\,Sabbi\\
 NGC\,5139 & Feb 14 -- Mar 24 2011 & 8$\times$800s & F275W & WFC3/UVIS & 12339 & E.\,Sabbi\\
 NGC\,5139 & Jul 15 2009 & 9$\times$350s+35s & F336W & WFC3/UVIS & 11452 & J.\,K.\,Quijano\\
 NGC\,5139 & Jan 10 -- Jul 04 2010 & 19$\times$350s & F336W & WFC3/UVIS & 11911 & E.\,Sabbi\\
 NGC\,5139 & Feb 14-15  2011 & 9$\times$350s & F336W & WFC3/UVIS & 12339 & E.\,Sabbi\\
 NGC\,5139 & Jul 26  2012 & 8$\times$700s$+$11$\times$10s & F336W & WFC3/UVIS & 12802 & J.\,MacKenty \\
 NGC\,5139 & Jul 15 2009 & 35s & F438W & WFC3/UVIS & 11452 & J.\,K.\,Quijano\\
 NGC\,5139 & Jan 14 -- Jul 04 2010 & 25$\times$438s & F438W & WFC3/UVIS & 11911 & E.\,Sabbi\\
 NGC\,5139 & Feb 15 -- Mar 24 2011 & 9$\times$350s & F438W & WFC3/UVIS & 12339 & E.\,Sabbi\\
 NGC\,5139 & Jul 15 2009 & 35s & F814W & WFC3/UVIS & 11452 & J.\,K.\,Quijano\\
 NGC\,5139 & Jan 10 -- Jul 04 2010 & 27$\times$40s & F814W & WFC3/UVIS & 11911 & E.\,Sabbi\\
 NGC\,5139 & Feb 15 -- Mar 24 2011 & 9$\times$40s & F814W & WFC3/UVIS & 12339 & E.\,Sabbi\\
 NGC\,5927 & Sep 01 2010 & 2$\times$475s$+$30s  & F336W & WFC3/UVIS & 11729 & W.\,Freedman \\
 NGC\,6341 & Oct 11 2010 & 2$\times$425s$+$30s  & F336W & WFC3/UVIS & 11729 & W.\,Freedman \\
 NGC\,6352 & Feb 13 2012 & 410s$+$5$\times$400s & F336W & WFC3/UVIS & 12746 & A.\,Kong \\
 NGC\,6362 & Aug 13 2010 & 5$\times$450s$+$368s & F336W & WFC3/UVIS & 12008 & A.\,Kong \\
 NGC\,6397 & Mar 9 -- 11 2010   & 6$\times$620s & F336W & WFC3/UVIS & 11633 & R.\,Rich \\
 NGC\,6535 & Sep 04 2010 & 5$\times$400s$+$253s & F336W & WFC3/UVIS & 12008 & A.\,Kong \\
 NGC\,6752 & May 05 2010 & 2$\times$500s$+$30s  & F336W & WFC3/UVIS & 11729 & W.\,Freedman \\
 NGC\,6752 & May 18 -- Sep 04 2011 & 12$\times$389s$+$6$\times$10s  & F435W & WFC/ACS & 12254 & A.\,Reiners \\

\hline\hline
\end{tabular}\\
\label{tab:data}
\end{table*}
%%%%%%%%%%%%%%%%%%%%%%%%%%%%%%%%%%%%%%%%%%%%%%%%%%%%%%%%%%%%%%%%%%%%%%%%%%      

\section{Multiple populations along the RGB}
\label{sec:map}

In the next subsections we  explain how we measured the intrinsic  $m_{\rm F275W}-m_{\rm F814W}$ and $C_{\rm F275W,F336W,F438W}$ RGB  width,  and used these two quantities to construct the chromosome map of each cluster.
 We then continue our analysis by using these maps to identify the 1G and 2G stellar populations along the RGB.

\subsection{The determination of the RGB color and pseudo-color  width}
\label{subsec:W}
%%%%%%%%%%%%%%%%%%%%%%%%%%%%%%%%%%%%% FIG 2 %%%%%%%%%%%%%%%%%%%%%%%%%%%%%%%%%%%
%
%
\begin{centering}
\begin{figure*}
 \includegraphics[width=13cm]{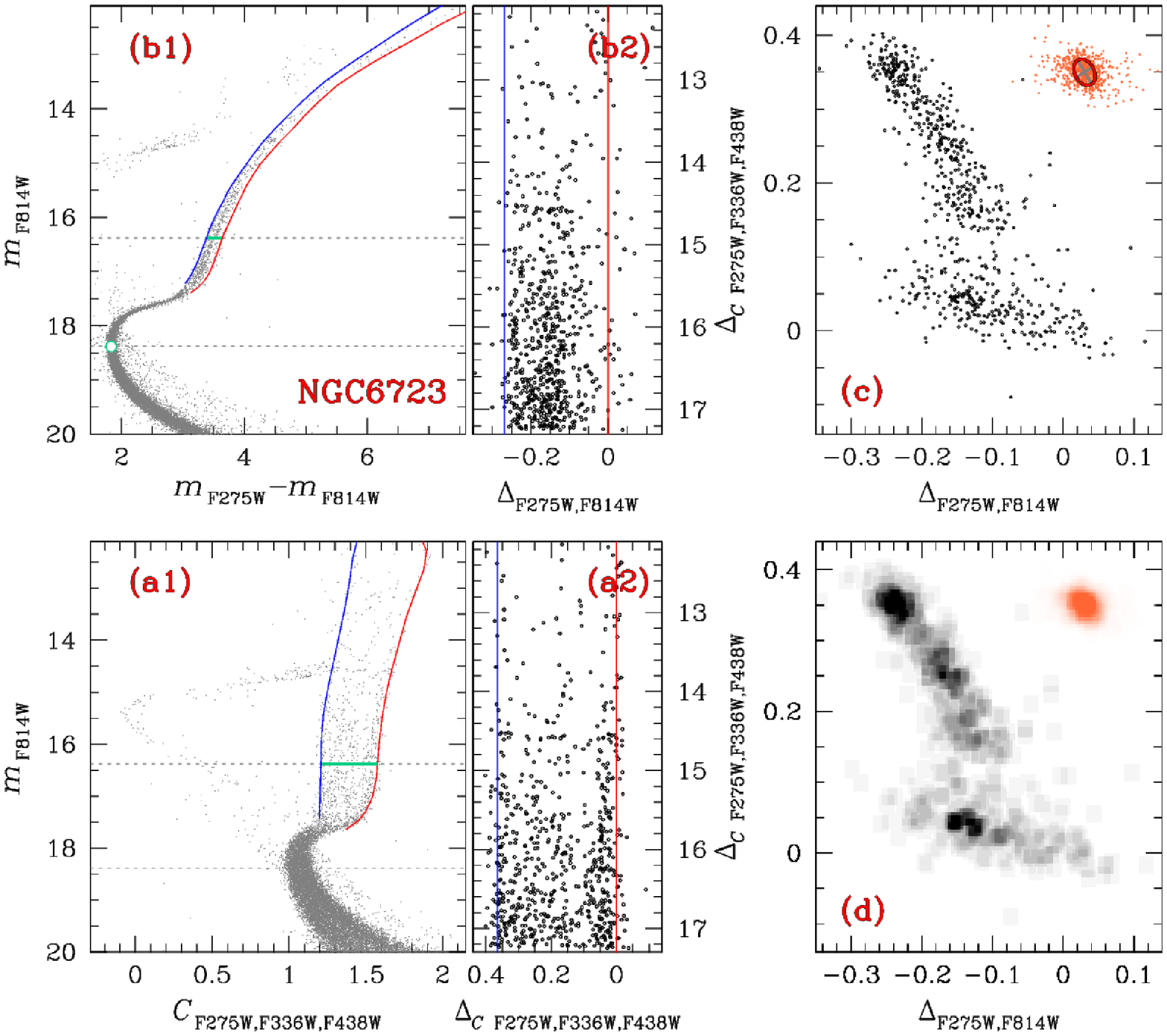}
%/home/milone/WORKS/treasury13/NGC6723/MATCH/popolazioni.macro figura
 \caption{This figure illustrates the procedure to derive the $\Delta_{\rm F275W,F336W,F438W}$ vs.\,$\Delta_{\rm F275W,F814W}$ pseudo two-color diagram (or `chromosome map')  for the prototypical cluster NGC\,6723. Panels (a1) and (b1) show the the $m_{\rm F814W}$ vs.\,$C_{\rm F275W,F336W,F438W}$ pseudo-CMD and the $m_{\rm F814W}$ vs.\,$m_{\rm F275W}-m_{\rm F814W}$ CMD  of NGC\,6723. The  aqua circle in panel (b1) marks the MS turn-off, whereas the two horizontal dotted lines in panels (a1) and (b1) are placed at the magnitude level of the MS turn-off and 2.0 F814W mag above it.  The blue and red lines mark the boundaries of the RGB, while the aqua segments in the panels (a1) and (b1) indicate the $m_{\rm F275W}-m_{\rm F814W}$ color and the $C_{\rm F275W,F336W,F438W}$ pseudo-color separation between the two lines at 2.0 F814W mag above the MS turn off. The `verticalized' $m_{\rm F814W}$ vs.\,$\Delta_{\rm C~ F275W,F336W,F438W}$ and $m_{\rm F814W}$ vs.\,$\Delta_{\rm F275W,F814W}$ diagrams for RGB stars are plotted in panels (a2) and (b2), respectively, where the red and blue vertical lines correspond to the RGB boundaries in panels (a1) and (b1) that translate into vertical lines in panel (a2) and (b2). 
The sample of RGB stars used to construct the chromosome map in panel (c) are those panels (a2) and (b2), where 
$\Delta_{\rm F275W,F336W,F438W}$ and \,$\Delta_{\rm C~ F275W,F814W}$ are defined in Equations (1) and (2) as explained in the text. The orange points indicate the distribution of stars expected from observational errors only, while the red ellipses includes the 68.27\% of the points. Panel (d) shows the Hess diagram for stars in panel (c). }
 \label{fig:metodoMappa}
\end{figure*}
\end{centering}
%%%%%%%%%%%%%%%%%%%%%%%%%%%%%%%%%%%%%%%%%%%%%%%%%%%%%%%%%%%%%%%%%%%%%%%%%%%%%%%

The color broadening of the RGB provides evidence for  the presence and diversity  of multiple stellar populations in GCs. Indeed, in a simple stellar population (made of chemically homogeneous and coeval stars)  the observed RGB width is entirely due to observational errors, whereas  the observed RGB width is much wider  than expected from  photometric errors   if multiple stellar populations are present. 

The procedure to estimate the RGB width in the $m_{\rm F814W}$ vs. $m_{\rm F275W}-m_{\rm F814W}$ and \,$C_{\rm F275W,F336W,F438W}$ plots  is illustrated in Figure~\ref{fig:metodoMappa} for the cluster NGC\,6723, taken as an example; the procedure is based in part on the naive estimator (Silverman 1986). We started by dividing the RGB into a series of F814W magnitude bins of size $\delta m$. The bins are defined over a grid of points separated by intervals  of fixed magnitude ($s=\delta m/3$). The procedure is extended to the RGB region fainter than  the HB level, where multiple sequences are more-clearly visible.

For each bin in F814W, we calculated the value of the 4$^{\rm th}$ and the 96$^{\rm th}$ percentile  of the $m_{\rm F275W}-m_{\rm F814W}$ and $C_{\rm F275W,F336W,F438W}$ distributions, to which we associated the mean F814W magnitude of RGB stars in each bin.
The resulting envelope of the RGB is represented by the red and blue lines in Figure~\ref{fig:metodoMappa}.  The smoothing has been performed by  boxcar averaging, where each point has been replaced by the average of the three adjacent points. Due to the small number of upper RGB stars, above the HB level, the  red and the blue envelopes in the region have been drawn by eye.

%%%%%%%%%%%%%%%%%%%%%%%%%%%%%%%%%%%%% FIG 2 %%%%%%%%%%%%%%%%%%%%%%%%%%%%%%%%%%%
\begin{centering}
\begin{figure*}
  \includegraphics[width=15cm]{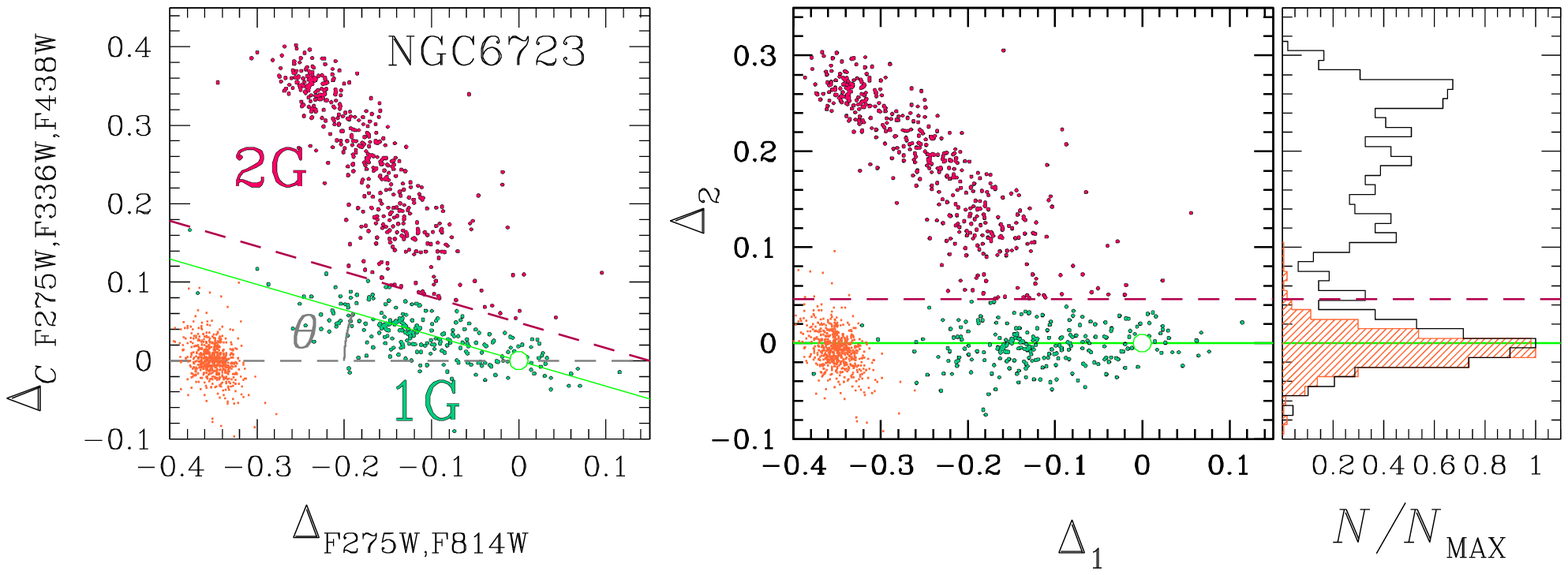}
  %/home/milone/WORKS/treasury13/mp2bin.macro figo
 \caption{
 The figure illustrates the method used to identify the two samples of bona-fide first generation (1G)  and second generation (2G)  stars in NGC\,6723.
 The left panel reproduces the $\Delta_{\rm F275W,F336W,F438W}$ vs.\,$\Delta_{\rm F275W,F814W}$ diagram from Figure~\ref{fig:metodoMappa}. 
 The green line through the origin of the frame is a fit to  the sequence of candidate 1G stars and defines an angle $\theta=18^\circ$ with respect to the horizontal line. The middle panel shows the $\Delta_{\rm 2}$ vs.\,$\Delta_{\rm 1}$ plot where these new coordinates have been obtained by  rotating  counterclockwise by an angle $\theta$  the plot in the left panel. The histogram in the right panel shows the distributions of the $\Delta_{\rm 2}$ values.  The orange points in the left and middle panels show the distribution of the observational errors and  their $\Delta_{\rm 2}$ distribution is represented by the shaded orange histogram in the right panel.
 The dashed magenta lines  separate the selected 1G and 2G stars, which are colored aqua and magenta, respectively, in the left and middle panels. See the text for details. 
 }
 \label{fig:p1p2}
\end{figure*}
\end{centering}
%%%%%%%%%%%%%%%%%%%%%%%%%%%%%%%%%%%%%%%%%%%%%%%%%%%%%%%%%%%%%%%%%%%%%%%%%%%%%%%

The observed RGB  width, $W_{\rm C~ F275W, F336W, F438W}^{\rm obs}$, has been derived as the difference between the $C_{\rm F275W,F336W,F438W}$ index of the red and blue fiducial line, calculated 2.0 F814W magnitudes above the MS turnoff, as illustrated in Panel (b1) of  Figure~\ref{fig:metodoMappa}. 
The error associated to $W_{\rm C~ F275W, F336W, F438W}^{\rm obs}$ has been determined by bootstrapping with replacements over  the sample of RGB stars, then repeated 1,000 times. The derived errors refer to one standard deviation  of the bootstrapped measurements.

The observed RGB width is partly intrinsic and  partly due to observational errors and limited statistics. The intrinsic RGB width, $W_{\rm C~ F275W, F336W, F438W}$, is calculated by subtracting in quadrature the errors affecting  the observed width,  which include both photometric errors and errors in the differential-reddening correction.
The same  procedure was adopted to measure  the intrinsic $m_{\rm F275W}-m_{\rm F814W}$ RGB color width, $W_{{\it m}_{\rm F275W}-{\it m}_{\rm F814W}}$, as illustrated in Panel (a1) of Figure~\ref{fig:metodoMappa}.
The results are listed in Table~\ref{tab:w} and reveal that for all the analyzed GCs,
 the RGB width is always significantly wider than expected from  the 
 errors alone, proving that all 57 GCs host multiple stellar populations.

\subsection{The `chromosome maps' of the multiple stellar populations }
\label{sec:mappe}
We now combine the pieces of information coming from both the $m_{\rm F814W}$ vs.\,$m_{\rm F275W}-m_{\rm F814W}$ color-magnitude diagram (CMD) and the $m_{\rm F814W}$ vs.\,$C_{\rm F275W,F336W,F438W}$ diagram to identify the multiple stellar populations in each GC.
To this end, we have used the  method illustrated in Figure~\ref{fig:metodoMappa}, analogous  to the technique introduced in Papers\,II and III,   and illustrated  here for the RGB of NGC\,6723.
Briefly,  we have `verticalized' the two diagrams  in such a way that the blue and the red fiducial lines translate into vertical lines. 
This is obtained by defining for each star:
\begin{equation}\label{eq:1}
\Delta_{\rm F275W,F814W}= W_{\rm F275W,F814W} \frac{X-X_{\rm fiducial~R}}{X_{\rm fiducial~R}-X_{\rm fiducial~B}}
\end{equation}
\begin{equation}\label{eq:2} 
\Delta_{ \rm C~ F275W,F336W,F438W}= W_{\rm C~ F275W,F336W,F438} \frac{Y_{\rm fiducial~R}-Y}{Y_{\rm fiducial~R}-Y_{\rm fiducial~B}}
\end{equation}
where $X$=($m_{\rm F275W}-m_{\rm F814W}$) and $Y$= $C_{\rm F275W,F336W,F438W}$ and `fiducial R' and `fiducial B' correspond to the red and the blue fiducial lines, respectively, as shown in panels (a2) and (b2) of Figure~\ref{fig:metodoMappa}. 
 
   Thus, $\Delta_{\rm F275W,F814W}=0$ and $\Delta_{\rm C~ F275W,F336W,F438W}= 0$ correspond to stars lying on the corresponding red fiducial line and the $\Delta$ quantities represent the color and pseudo-color distance from such lines.
The resulting $\Delta_{\rm C~ F275W,F336W,F438W}$ vs.\,$\Delta_{\rm F275W,F814W}$ plot is shown in panels (c) and (d) and reveals the distinct stellar populations of NGC\,6723.

Following the nomenclature introduced in Paper V, we will refer to plots such as those shown in Panels (c) and (d) of Figure~\ref{fig:metodoMappa}   as 
the `chromosome map' of a GC.
The chromosome maps of all 57 GCs are presented in Section~~\ref{sec:rea}.

\subsection{Distinguishing First and Second Generation stars}
\label{sub:G1G2}

The chromosome map of NGC\,6723 shown in Panels (c) and (d) of Figure~\ref{fig:metodoMappa} and in the left panel of Figure~\ref{fig:p1p2} reveals that cluster stars are distributed along two main, distinct groups that we name 1G and 2G and that correspond to the first and second generation of stars as defined in Paper V.
It is indeed commonly believed that the multiple stellar populations phenomenon in GCs
is the result of multiple events of star formation where 2G stars form out  of material processed by 1G stars (e.g., Decressin et al.\,2007, Paper V, D'Antona et al.\,2016 and references therein). Thus, in this paper we will consider GC `multiple populations' and `multiple generations' as synonyms, as  done in previous papers of this series.

We preliminarily identify 1G stars as those at nearly constant 
  $\Delta_{\rm C~F275W,F336W,F438W}$ departing from  the origin of the reference frame, located at $\Delta_{\rm C~F275W,F336W,F438W} = \Delta_{\rm F275W,F814W} =0$.  As a consequence, 2G stars are identified as those in the steep branch reaching high values of $\Delta_{\rm C~ F275W,F336W,F438W}$.
A full justification of this choice is presented in Section~\ref{sec:chemistry}, where 1G and 2G are chemically tagged, in analogy to what done in Paper II and Paper III. 

The procedure to define a sample of bona-fide 1G and 2G stars is illustrated in Figure~\ref{fig:p1p2} for NGC\,6723. The green line is a fit to the group of 1G stars
and the  angle between the green line and the dashed horizontal line is $\theta=18^{\rm o}$. We adopted this same value of $\theta$ for all the analyzed clusters.
The $\Delta_{\rm 2}$ versus  $\Delta_{\rm 1}$ diagram shown in the middle panel of Figure~\ref{fig:p1p2} has been obtained by rotating  counterclockwise the left-panel diagram by an angle $\theta$ around the origin of the reference frame, and the black histogram plotted in the right panel  represents the normalized $\Delta_{\rm 2}$ distribution of cluster stars. The orange points shown in the left and middle panels of Figure~\ref{fig:p1p2} represent the expected distribution of the observational errors obtained by Montecarlo simulations and have been plotted at the arbitrary  position $\Delta_{\rm 2}=0$.
The normalized histogram distribution of the $\Delta_{\rm 2}$  errors is shown in orange in the right panel of the figure.  The magenta dashed line is then plotted at the $\Delta_{\rm 2}$ level corresponding to the $3\sigma$ deviation from the mean of the error histogram, and the same line is also reported in the left panel, after counter-rotation. 

We have then taken  as bona fide 1G stars all those below the magenta dashed line, while the remaining  stars are defined as 2G. 1G and 2G stars are colored aqua and magenta, respectively, in the left  and middle panel of Figure~\ref{fig:p1p2}.
We can already notice that the $\Delta_{\rm F275W,F814W}$ and $\Delta_{\rm C~ F275W,F336W,F438W}$ extension of both 1G and 2G stars in this cluster  is significantly wider than expected from photometric errors alone, thus demonstrating that both 1G and 2G stars  in the cluster are not chemically homogeneous. 
As we shall see, this is the case for the vast majority of our 57 GCs.

\section{The chromosome maps of the 57 globular clusters}
\label{sec:rea}
%%%%%%%%%%%%%%%%%%%%%%%%%%%%%%%%%%%%% FIG 2 %%%%%%%%%%%%%%%%%%%%%%%%%%%%%%%%%%%
\begin{centering}
\begin{figure*}
 \includegraphics[width=15cm]{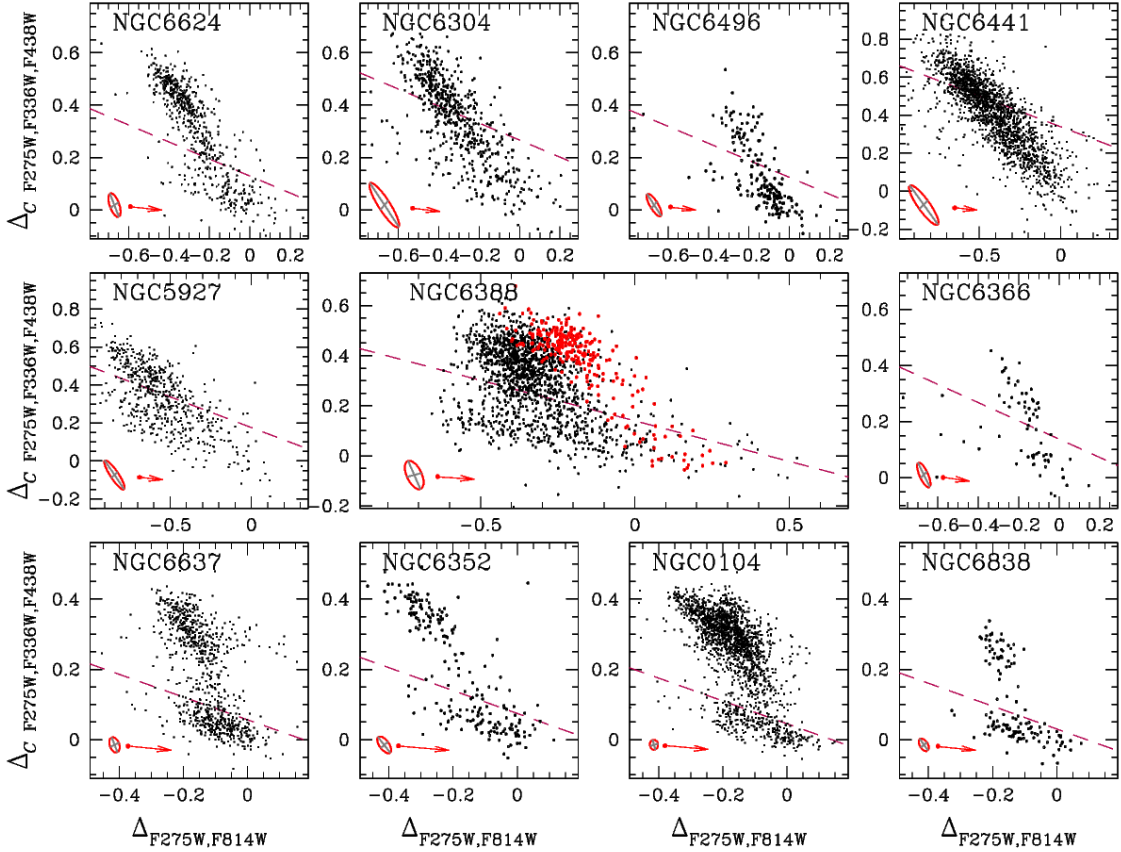}
%/home/milone/WORKS/treasury13/mp2bin.macro
 \caption{ $\Delta_{\rm C~ F275W,F336W,F438W}$ vs.\,$\Delta_{\rm F275W,F814W}$ diagrams, or chromosome maps, for RGB stars in 11 GCs. Namely NGC\,6624, NGC\,6304, NGC\,6496, NGC\,6441, NGC\,5927, NGC\,6388, NGC\,6366, NGC\,6637, NGC\,6352, NGC\,104 (47\,Tucanae), and NGC\,6838 (M\,71). Clusters are approximately sorted according to their metallicity, from the most metal rich, to the most metal poor. The ellipses are indicative of the observational errors and include 68.27\% of the simulated stars. The magenta dashed line is used to separate bona-fide 1G from 2G stars and has been determined as in Section~\ref{sub:G1G2}. Red points indicate red-RGB stars and will be selected and discussed in Section~\ref{sec:ano}, while the arrows indicate the reddening vector and correspond to a reddening variation of $\Delta E(B-V$)=0.05. Note, however, that all these plots are constructed using photometric data corrected for differential reddening. } 
 \label{fig:maps1}
\end{figure*}
\end{centering}
%%%%%%%%%%%%%%%%%%%%%%%%%%%%%%%%%%%%%%%%%%%%%%%%%%%%%%%%%%%%%%%%%%%%%%%%%%%%%%%

%%%%%%%%%%%%%%%%%%%%%%%%%%%%%%%%%%%%% FIG 2 %%%%%%%%%%%%%%%%%%%%%%%%%%%%%%%%%%%
\begin{centering}
\begin{figure*}
 \includegraphics[width=15cm]{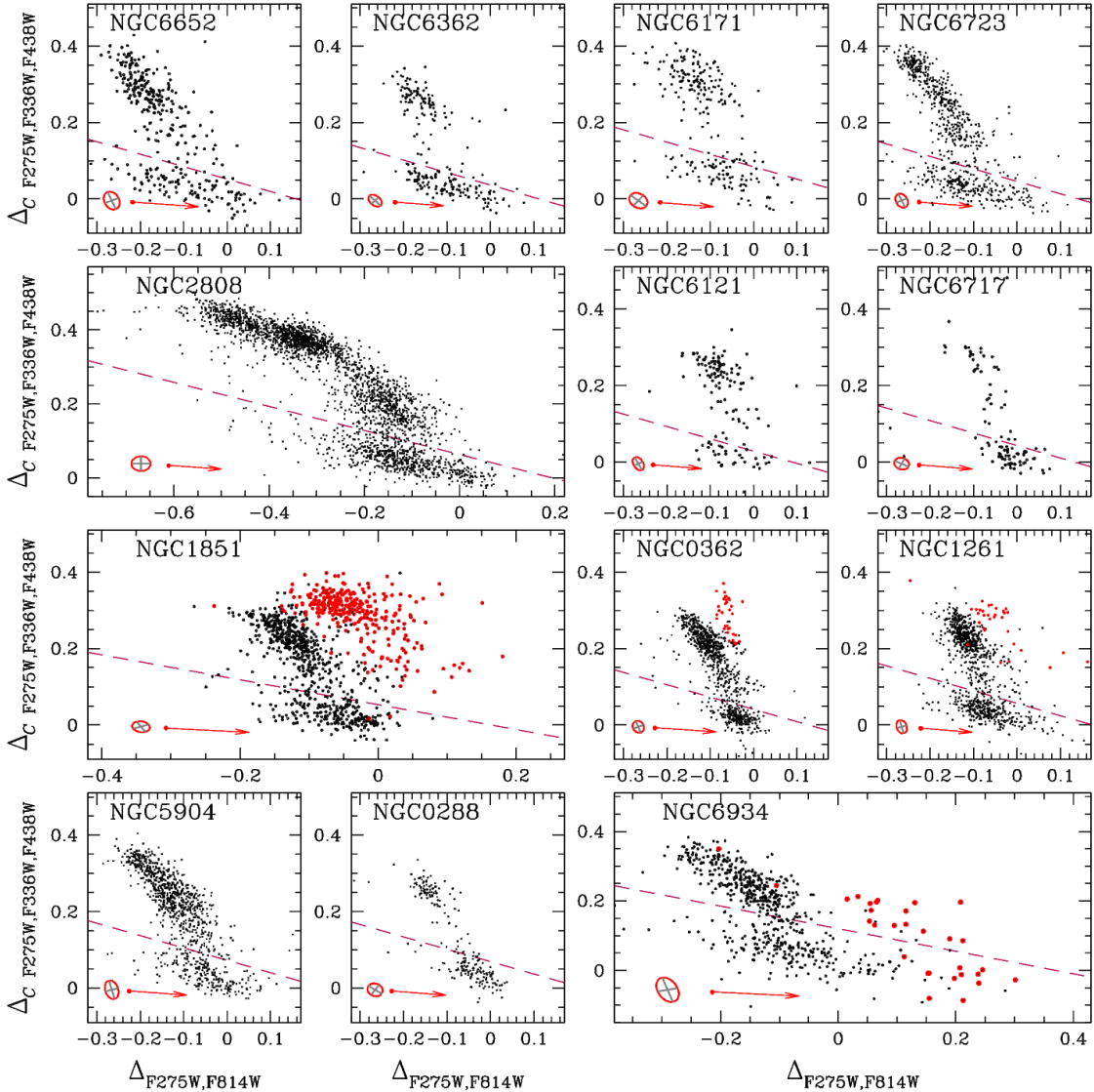}
 \caption{As in Figure~\ref{fig:maps1}, but for NGC\,6652, NGC\,6362, NGC\,6171 (M\,107), NGC\,6723,  NGC\,2808, NGC\,6121 (M\,4), NGC\,6717, NGC\,1851, NGC\,362, NGC\,1261, NGC\,5904 (M\,5), NGC\,288, and NGC\,6934. }
 \label{fig:maps2}
\end{figure*}
\end{centering}
%%%%%%%%%%%%%%%%%%%%%%%%%%%%%%%%%%%%%%%%%%%%%%%%%%%%%%%%%%%%%%%%%%%%%%%%%%%%%%%

%%%%%%%%%%%%%%%%%%%%%%%%%%%%%%%%%%%%% FIG 2 %%%%%%%%%%%%%%%%%%%%%%%%%%%%%%%%%%%
\begin{centering}
\begin{figure*}
 \includegraphics[width=15cm]{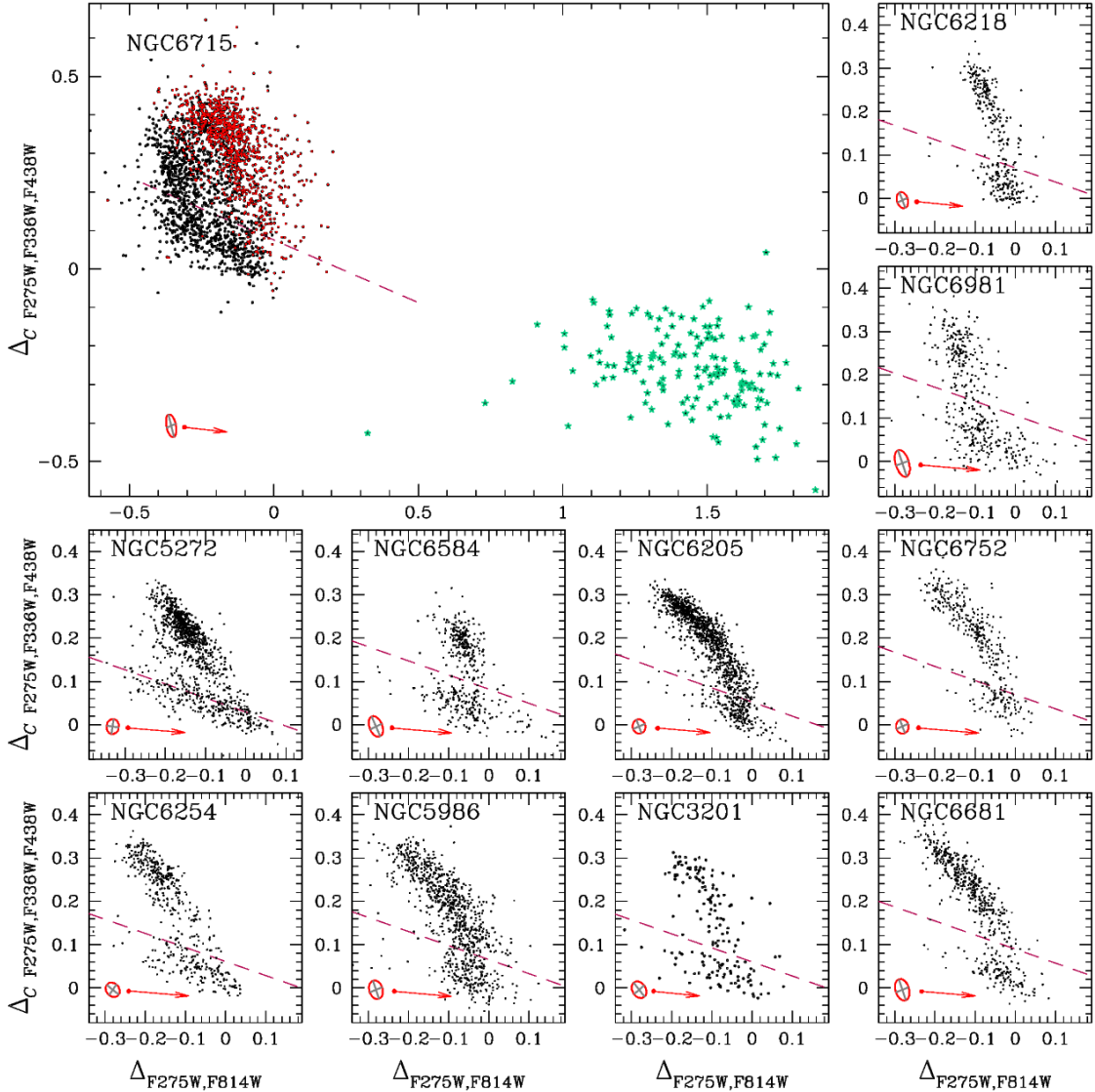}
 \caption{As in Figure~\ref{fig:maps1},  but for the stellar system formed by NGC\,6715 (M\,54),  and for NGC\,6218 (M\,12), NGC\,6981 (M\,72), NGC\,5272 (M\,3), NGC\,6584, NGC\,6205 (M\,13), NGC\,6752, NGC\,6254 (M\,10), NGC\,5986, NGC\,3201, and NGC\,6681 (M\,70).  The aqua starred symbols in the map of M\,54 indicate stars of the metal rich population in the core of the Sagittarius dwarf galaxy, to which M54 belongs.}
 \label{fig:maps3}
\end{figure*}
\end{centering}
%%%%%%%%%%%%%%%%%%%%%%%%%%%%%%%%%%%%%%%%%%%%%%%%%%%%%%%%%%%%%%%%%%%%%%%%%%%%%%%

%%%%%%%%%%%%%%%%%%%%%%%%%%%%%%%%%%%%% FIG 2 %%%%%%%%%%%%%%%%%%%%%%%%%%%%%%%%%%%
\begin{centering}
\begin{figure*}
 \includegraphics[width=15cm]{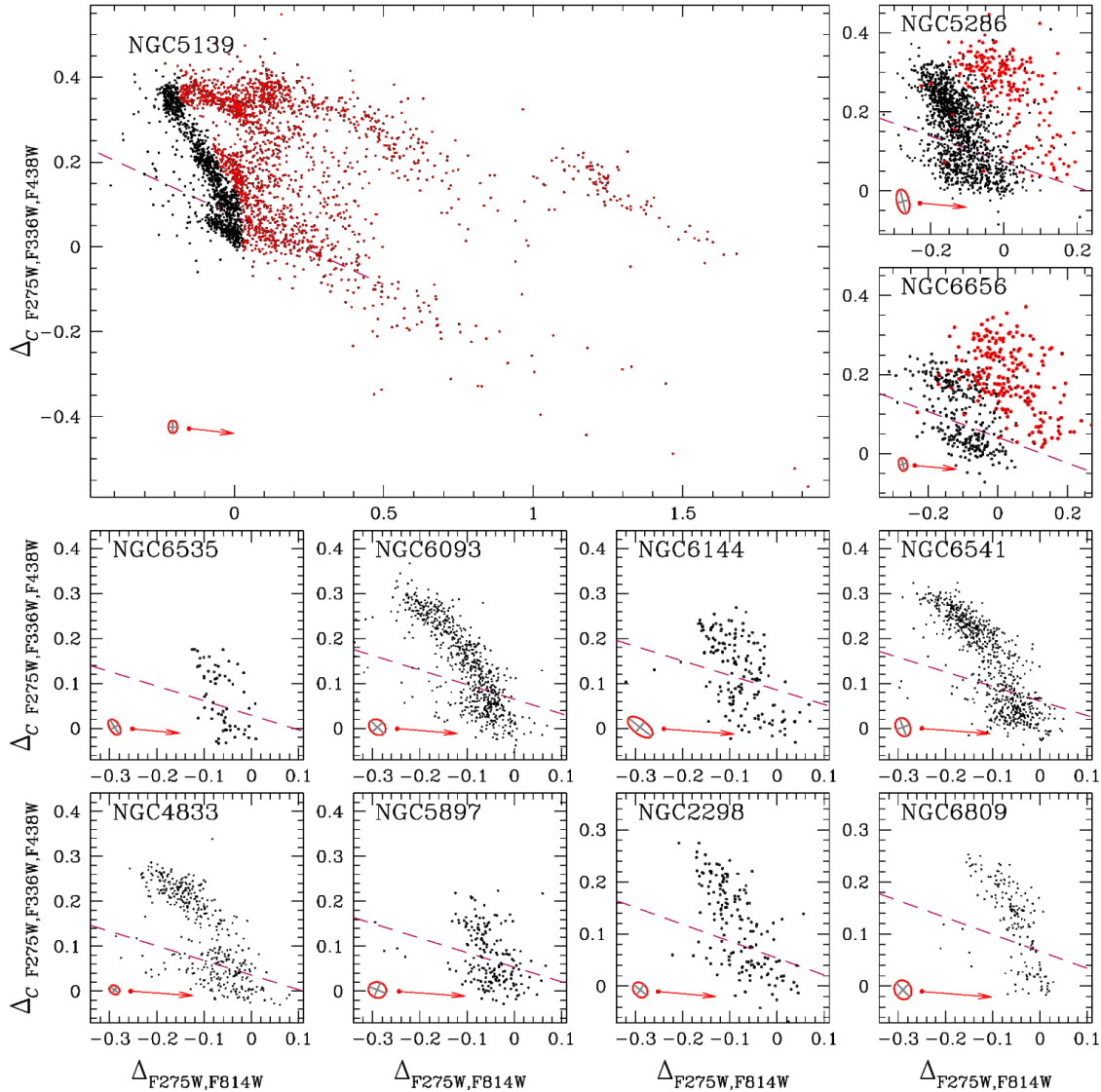}
 \caption{As in Figure~\ref{fig:maps1},  but for NGC\,5139 ($\omega$ Centauri), NGC\,5286, and NGC\,6656 (M\,22), NGC\,6535, NGC\,6093 (M\,80), NGC\,6144, NGC\,6541, NGC\,4833, NGC\,5897, NGC\,2298, and NGC\,6809 (M\,55).}
 \label{fig:maps4}
\end{figure*}
\end{centering}
%%%%%%%%%%%%%%%%%%%%%%%%%%%%%%%%%%%%%%%%%%%%%%%%%%%%%%%%%%%%%%%%%%%%%%%%%%%%%%%

%%%%%%%%%%%%%%%%%%%%%%%%%%%%%%%%%%%%% FIG 2 %%%%%%%%%%%%%%%%%%%%%%%%%%%%%%%%%%%
\begin{centering}
\begin{figure*}
 \includegraphics[width=15cm]{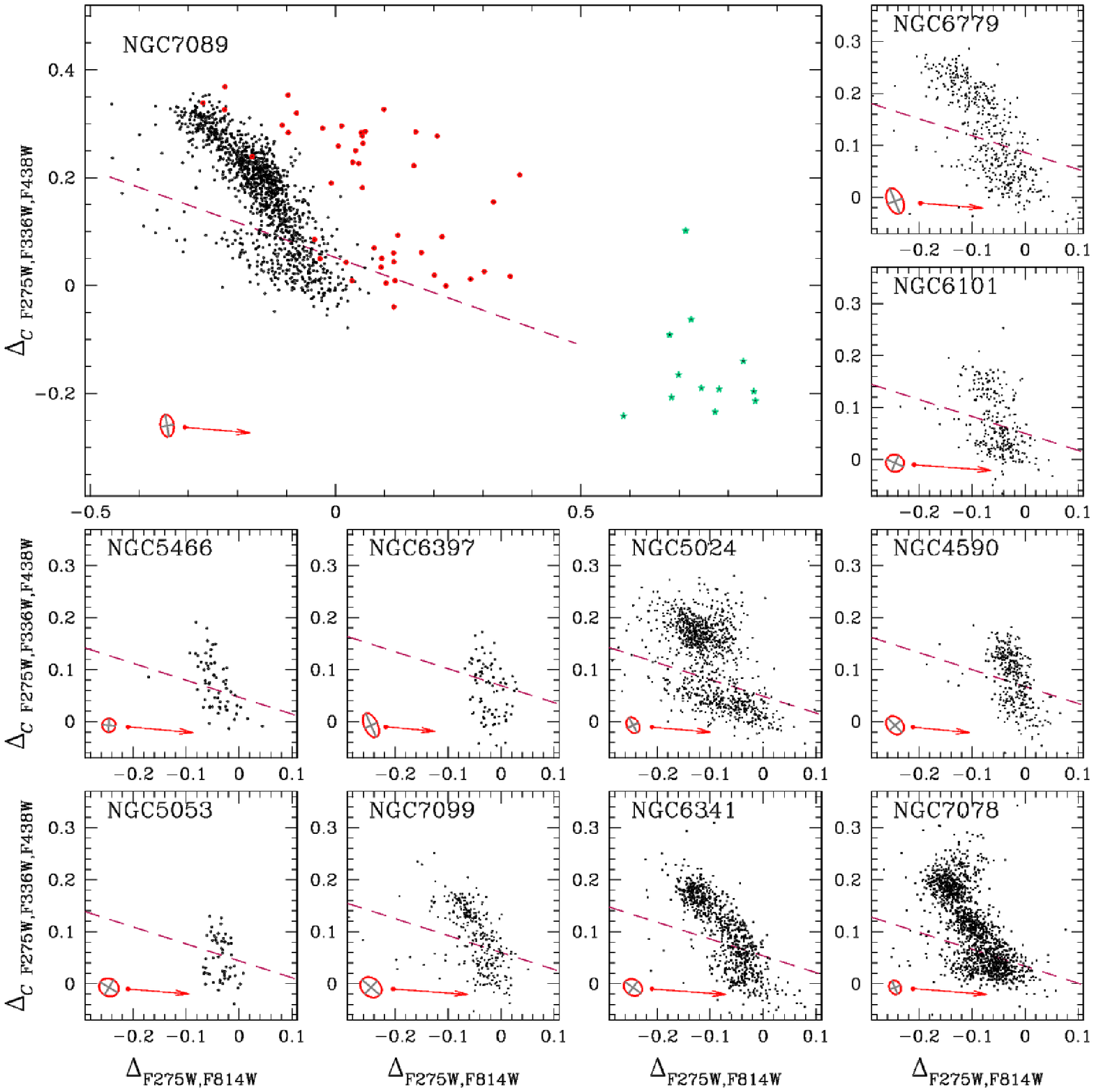}
 \caption{As in Figure~\ref{fig:maps1}, but for NGC\,7089 (M\,2), NGC\,6779 (M\,56), NGC\,6101, NGC\,5466, NGC\,6397, NGC\,5024, NGC\,4590, NGC\,5053, NGC\,7099 (M\,30), NGC\,6341 (M\,92), and NGC\,7078 (M\,15). Stars in the most metal rich population of NGC\,7089 are represented with aqua starred symbols.}
 \label{fig:maps5}
\end{figure*}
\end{centering}
%%%%%%%%%%%%%%%%%%%%%%%%%%%%%%%%%%%%%%%%%%%%%%%%%%%%%%%%%%%%%%%%%%%%%%%%%%%%%%%

Figures~\ref{fig:maps1} to~\ref{fig:maps5} show a collection of the chromosome maps for all 57 GCs studied in this paper. 
GCs are roughly sorted in order of decreasing metallicity, from the most metal rich (NGC\,6624, [Fe/H]=$-$0.44, Figure~\ref{fig:maps1}) to the most metal poor (NGC\,7078, [Fe/H]=$-$2.37, Figure~\ref{fig:maps5}).

\subsection{Classifying clusters in two main types}

In most maps it is possible to easily identify the two main groups of 1G and 2G stars  as it was the case for  NGC\,6723 (see Figure~\ref{fig:p1p2}).  The magenta dashed lines superimposed on each panel of Figures~\ref{fig:maps1}--~\ref{fig:maps5} have been derived as described in Section~\ref{sub:G1G2} and have been used to identify the two groups of bona fide 1G and 2G stars of each cluster. Clusters for which the map allows the 1G/2G distinction as described for NGC 6723 are called here Type I clusters.  However, the extension of the 1G group of stars and its separation from the 2G group are quite ambiguous in some clusters, and eventually a distinction between 1G and 2G groups is no longer possible, at least with the present photometric accuracy. This is the case for the three clusters  NGC\,5927, NGC\,6304, and NGC\,6441.  The 1G/2G separation may still be possible using other passbands, such as in the case of NGC\,6441 (Bellini et al.\,2013).

 Finally,  several other clusters exhibit more complex 
chromosome maps, with an additional 2G sequence (e.\,g.\,NGC\,1851) or even what appears to be a split of both 1G and 2G sequences (e.\,g.\,NGC\,6934). Stars in these additional sequences are colored in red in the chromosome maps.  These are the clusters that define the Type II and, besides the mentioned NGC\,1851 and NGC\,6934, this group includes NGC\,362, NGC\,1261, NGC\,5286,  NGC\,6388, NGC\,6656, NGC\,6715, NGC\,7089 and the famous $\omega$ Cen which, not surprisingly, has the most complex map of them all.
 Noticeably, in order to derive the red and blue fiducial lines that are used to determine the chromosome map of Type-II GCs (see Figure~\ref{fig:metodoMappa}), we used only blue-RGB stars.
Type II clusters deserve a dedicated analysis, which is presented in Section~\ref{sec:ano}.  

As illustrated by Figures~\ref{fig:maps1} to \ref{fig:maps5}, the chromosome maps of of Type I  GC exhibit a great deal of variety. In particular, the $\Delta_{\rm C~F275W,F336W,F438W}$ and  $\Delta_{\rm F275W,F814W}$ extensions differ  from one cluster to another, and in several clusters distinct clumps are clearly visible within the 1G and/or the 2G sequences. This is the case of  NGC\,2808 where at least five distinct sub-populations can be identified, as already illustrated in Paper III. 
The detailed study of substructures within the 1G and 2G  sequences is not further developed in the present paper.
 
 Among Type I clusters, quite surprising is the case of NGC\,6441, often considered a twin cluster of NGC\,6388, since both are metal rich clusters with an extended blue HB (e.g., Rich et al.\,1997; Bellini et al.\,2013, and references therein). Yet, their chromosome maps are radically different, with the Type II NGC\,6388 exhibiting a very complex map whereas the Type I NGC\,6441 shows a unique sequence where it is   not even possible to distinguish between 1G and 2G stars. Similarly, we note significant difference between the chromosome map of the second-parameter pair cluster NGC\,6205 (M\,13) and NGC\,5272 (M\,3), with the latter hosting a very extended 1G.  First- and second-generation stars in the other famous second-parameter pair, NGC\,288 NGC\,362, share a similar distribution in the corresponding chromosome maps. Intriguingly, NGC\,362 hosts a poorly-populated red-RGB, which is not present in NGC\,288.
 
\subsection{The fraction of 1G stars}
\label{sub:G1G2}

\begin{centering}
\begin{figure}
 \includegraphics[width=8.7cm]{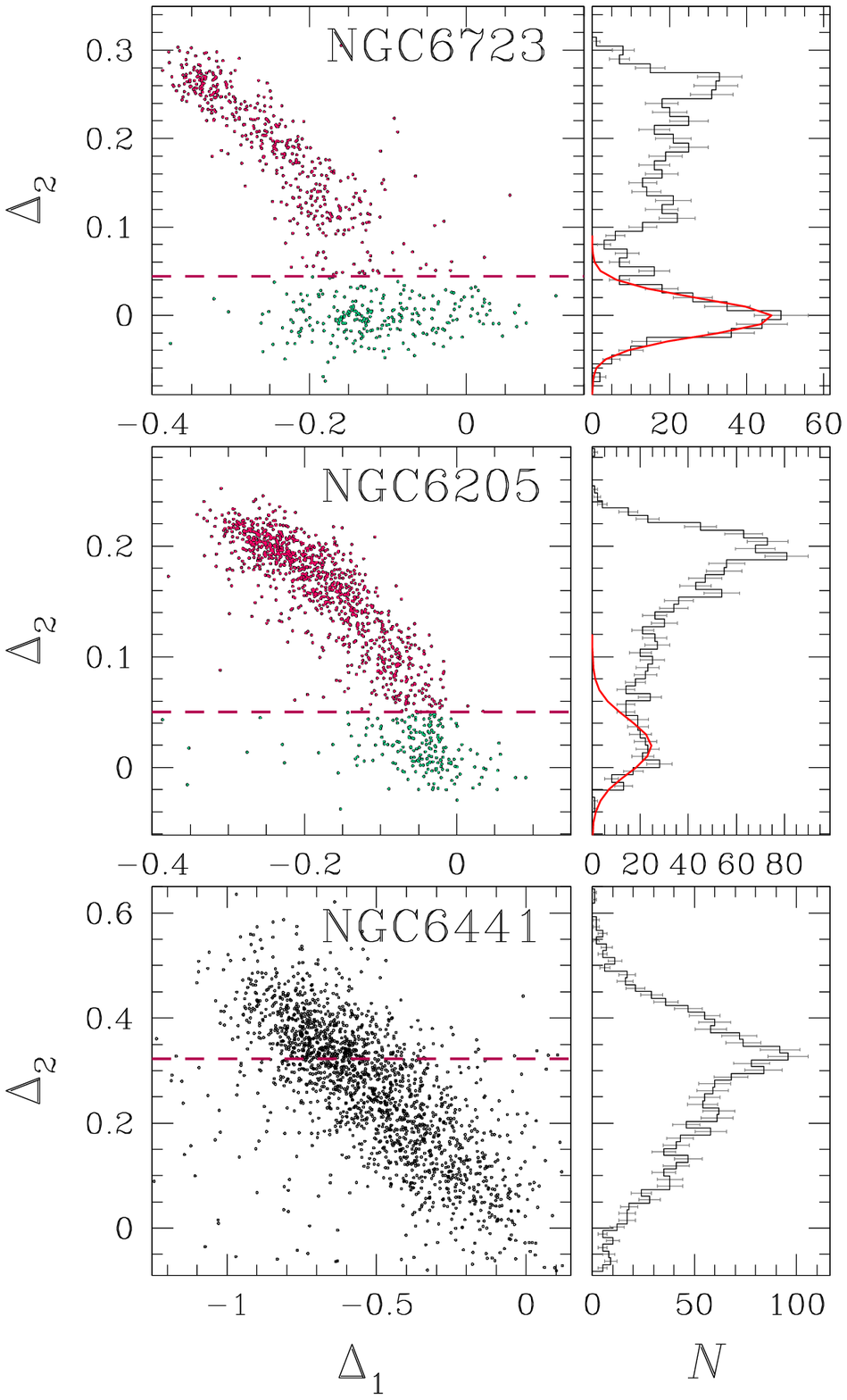}
%/home/milone/WORKS/treasury13/mp2bin.macro pione
 \caption{This figure exemplifies the procedure to estimate the fraction of 1G stars with respect to the total number of RGB stars,  for NGC\,6723 (upper panels) and NGC\,6205 (middle panels). 
   The left panels show the $\Delta_{2}$ vs.\,$\Delta_{1}$ diagrams presented in Section~\ref{sub:G1G2}, where the pre-selected 1G and 2G stars are colored aqua and magenta, respectively. The histogram in the right panels show the distributions of the  $\Delta_{\rm 2}$ values . The red  lines superimposed on the histograms of NGC\,6723 and NGC\,6205 are the best-fitting  Gaussians of the 1G peak of the histogram. The fraction of 1G stars is then calculated as the ratio of the area of the Gaussian over that of the whole histogram.
Lower panels show the case of NGC\,6441, a Type I cluster, for which no clear distinction can be made between 1G and 2G stars and, correspondingly, we  did not estimate the fraction of 1G stars.   See the text for details. }
 \label{fig:PRschema}
\end{figure}
\end{centering}

The procedure to estimate the fraction of 1G-stars with respect to the total number of studied RGB stars (${\rm N}_{\rm TOT}$) is illustrated in the upper panels of Figure~\ref{fig:PRschema} for NGC\,6723, where we reproduce the $\Delta_{\rm 2}$ versus  $\Delta_{\rm 1}$ plot shown in Figure~\ref{fig:p1p2}, now having  colored 1G and 2G stars aqua and magenta, respectively. The corresponding histogram distribution of  $\Delta_{\rm 2}$  is plotted in the upper-right panel of Figure~\ref{fig:PRschema}.  The Gaussian fit to the distribution of bona fide 1G stars selected in Section~\ref{sec:mappe} is  represented by the red continuous line. The fraction of 1G stars (${\rm N}_{1}/{\rm N}_{\rm TOT}$) has been derived as the ratio between the area under the Gaussian and the total number of RGB stars in the chromosome map.

The middle panels of Figure~\ref{fig:PRschema}  illustrate the procedure described above, now applied to NGC\,6205 where the separation between 1G and 2G stars is much less evident than for NGC\,6723. NGC\,6205 is the most uncertain case for a cluster that we classified as Type I. The lower panels of  Figure~\ref{fig:PRschema} show the case for NGC\,6441, where there is no appreciable separation between  1G and 2G stars, making NGC\,6441 a typical example of a Type I cluster for which we did not attempt to estimate the fraction of  1G stars. 

 The derived fractions of 1G stars are listed in Table~\ref{tab:w}  which also provides the total number of RGB stars included in the chromosome map and the ratio between the maximum radius of the analyzed stars and the cluster half light radius. Radial gradients in the distribution of the 1G and 2G stars are indeed known to exist in some clusters (e.g., Sollima et al.\,2007; Bellini et al.\,2009, 2013; Milone et al. 2012b; Cordero et al.\,2014;  Johnson \& Pilachowski 2010) hence this ratio provides a rough indication of the relative number of stars within the analyzed field of view with respect to the entire cluster stellar population.
 
%%%%%%%%%%%%%%%%%%%%%%%%%%%%%%%%%%%%% FIG 2 %%%%%%%%%%%%%%%%%%%%%%%%%%%%%%%%%%%
%%%%%%%%%%%%%%%%%%%%%%%%%%%%%%%%%%%%%%%%%%%%%%%%%%%%%%%%%%%%%%%%%%%%%%%%%%%%%%%

A visual inspection of the maps shown in Figures~\ref{fig:maps1}~--~\ref{fig:maps5} reveals that the $\Delta_{\rm F275W, F814W}$ extension of 1G- and 2G-stars may  dramatically  differ from one cluster to another. For example, in NGC\,6205 and NGC\,6752 the second generation is significantly more extended than the first one, while in NGC\,5024 and NGC\,5272 1G and 2G stars have a similar extension.

 In order to quantify the $\Delta_{\rm F275W, F814W}$ extension of 1G and 2G stars, we determined the width of the 1(2)G, $W^{\rm obs, 1(2)G}_{\rm F275W, F814W}$, as the difference between the 90$^{\rm th}$ and the 10$^{\rm th}$ percentile of the $\Delta_{\rm F275W, F814W}$ distribution of 1(2)G stars. 
 The intrinsic width has been estimated by subtracting the color errors in quadrature (including errors associated to the differential reddening corrections).
 The  values of $W^{\rm 1(2)G}_{\rm F275W, F814W}$ are also listed in Table~\ref{tab:w}. 

 As already noted, the fact that $W^{\rm 1(2)G}_{\rm F275W, F814W}$ is significantly larger than zero in most GCs, (i.e., the observed 1G and 2G widths are larger than measurement errors) demonstrates that neither 1G nor 2G  are consistent with a simple stellar population. This  raises a new fundamental question: what are the chemical differences within the 1G population of a GC?

\subsection{The chemical composition of multiple stellar populations}
\label{sec:chemistry}
The chemical characterization of the multiple populations identified on  the chromosome maps  is a key step to justify our identification of 1G and 2G stars 
as belonging to the first and the second generation and an indispensable tool to understand their origin.  For this purpose  the spectroscopic analysis of some stars included in our chromosome maps is needed. At present, we
can rely only on existing data but additional extensive spectroscopic surveys are needed to shed further light on our photometric data.

\begin{centering}
\begin{figure}
%/home/milone/WORKS/treasury13/NGC5286/MATCH/spectroscopy/macro go gob
 \includegraphics[width=8cm]{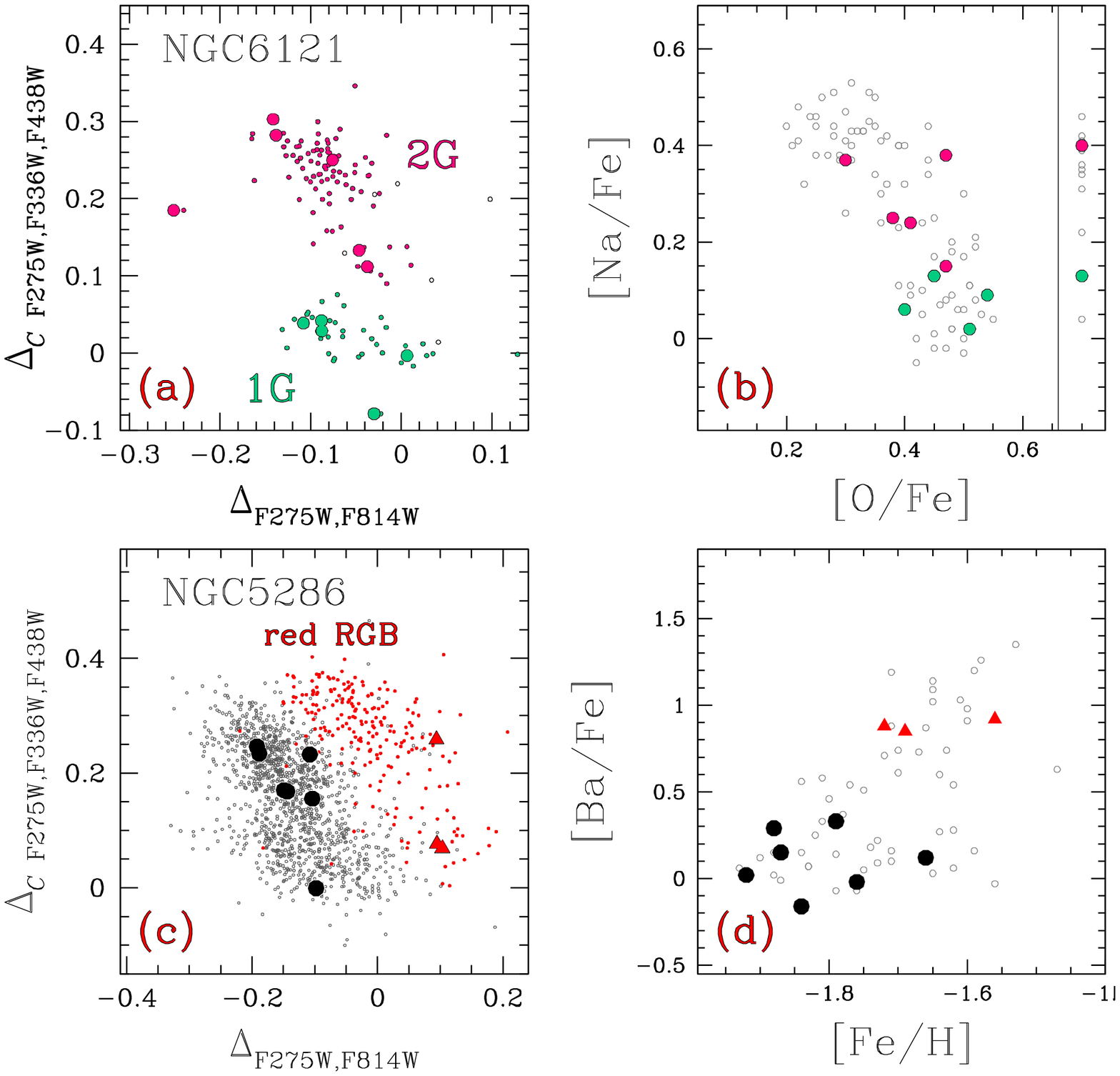}
 \caption{ Panel (a) shows the chromosome map of RGB stars in the Type I cluster NGC\,6121, where we have colored aqua and magenta 1G and 2G stars, respectively. Large aqua and magenta dots indicate 1G and 2G stars studied spectroscopically by Marino et al.\,(2008), and whose  [Na/Fe] vs.\,[O/Fe] anticorrelation is shown in panel (b) using the same symbols. Stars for which an oxygen abundance determination is  not available are plotted on the right side of the vertical line in this and in similar panels.
   The chromosome map of the Type II NGC\,5286 is shown in panel (c), where red-RGB stars are colored red and black point are used for  the remaining RGB stars.   Large black circles and red triangles indicate those stars studied spectroscopically by Marino et al.\,(2015), and whose [Ba/Fe]  vs.\,[Fe/H] plot is shown in panel (d).}
 \label{fig:spettri}
\end{figure}
\end{centering}

To illustrate the case,  in Figure~\ref{fig:spettri} we focus on NGC\,6121 as a
prototype of a Type I cluster. 
Multiple stellar populations in NGC\,6121 have been widely studied, both photometrically (e.g.\,Marino et al.\,2008; Milone et al.\,2014; Nardiello et al.\,2015b) and spectroscopically (e.g.\,Ivans et al.\,1999; Marino et al.\,2008, 2011; Carretta et al.\,2009, 2013). 
From Marino et al.\,(2008), chemical analysis is available for eleven stars  in common with our WFC3/UVIS sample of RGB  stars.
Panel (b) of  Figure~\ref{fig:spettri} shows  the sodium-oxygen anticorrelation, where some  stars are oxygen rich and sodium poor, hence  with primordial chemical composition,  while  others are  enhanced in sodium and depleted in oxygen.
These stars are shown, respectively, with aqua and magenta filled circles in  panels (a) and (b) of Figure~\ref{fig:spettri}, showing  that  those we have called 1G stars have indeed primordial chemical composition, while  2G stars are Na-rich and O-poor. No significant differences  in iron content appear to exist among 1G and 2G stars in NGC\,6121. In Papers II and III we performed a similar chemical tagging   for NGC\,7089 and for NGC\,2808, by comparing the chromosome map of these clusters with the light-element abundances from Yong et al.\,(2014) and Carretta et al.\,(2006), respectively.

The chemical tagging of stars identified on the chromosome maps is clearly very limited at this time, but it could be greatly expanded by future spectroscopic observations targeting stars selected on the chromosome maps illustrated in this paper.
The other panels in Figure~\ref{fig:spettri} refer to the Type-II GC NGC\,5286 and will be used in the next Section dedicated to Type II clusters.

%%%%%%%%%%%%%%%%%%%%%%%%%%%%%%%%%%%%% FIG 2 %%%%%%%%%%%%%%%%%%%%%%%%%%%%%%%%%%%
%%%%%%%%%%%%%%%%%%%%%%%%%%%%%%%%%%%%%%%%%%%%%%%%%%%%%%%%%%%%%%%%%%%%%%%%%%%%%%%

%%%%%%%%%%%%%%%%%%%%%%%%%%%%%%%%%%%%% FIG 2 %%%%%%%%%%%%%%%%%%%%%%%%%%%%%%%%%%%
\begin{centering}
\begin{figure}
%/home/milone/WORKS/treasury13/NGC5286/MATCH/spectroscopy/macro go gob
 \includegraphics[width=8cm]{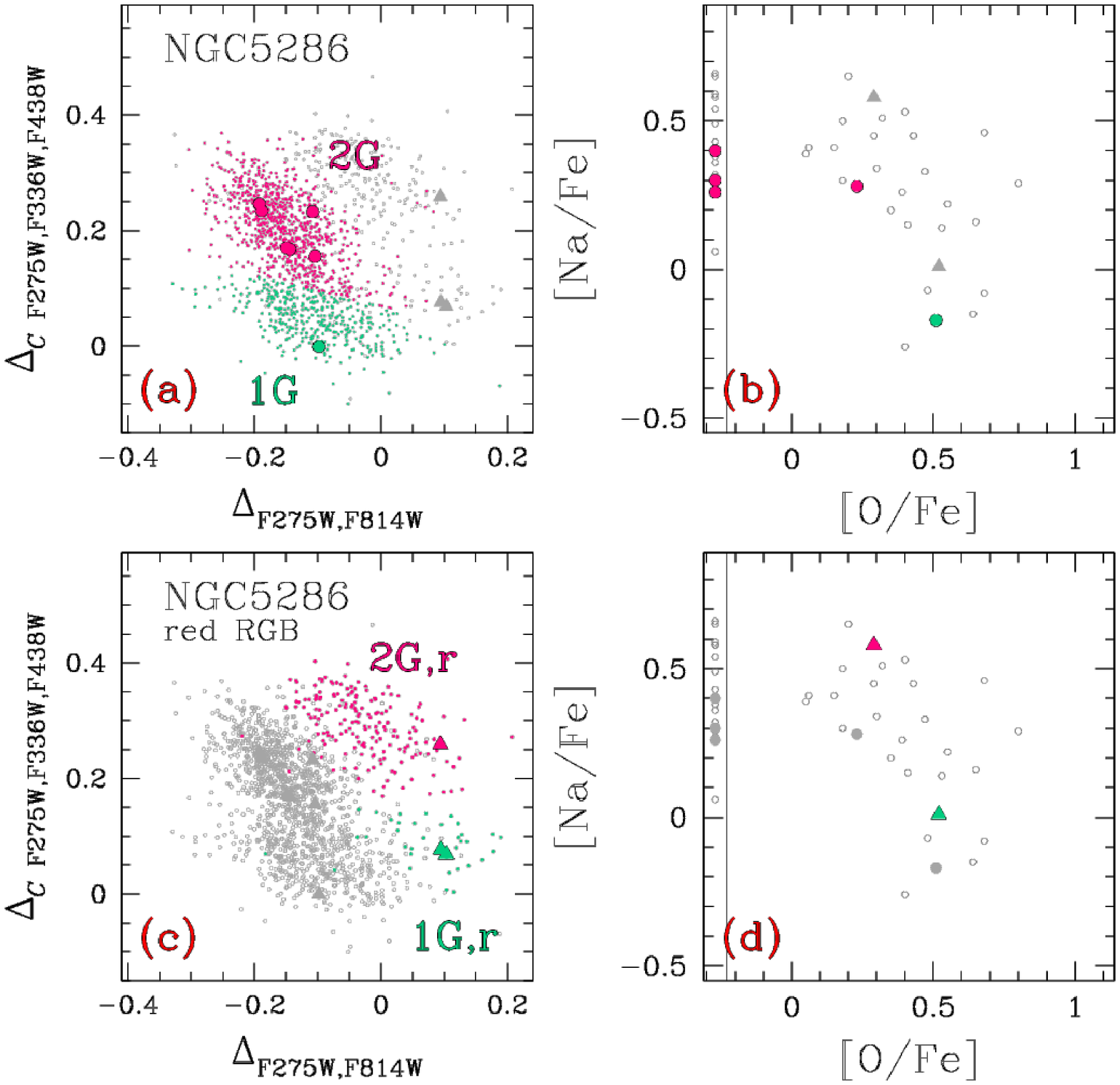}
 \caption{ 
   In the chromosome map of NGC\,5286, shown in panel (a) only stars from the blue-RGB are used, i.e., those colored black in Figure~\ref{fig:SGB5286}c. Aqua and magenta colors highlight 1G and 2G stars, respectively, with stars also studied spectroscopically are represented by large filled symbols and whose  [Na/Fe] vs.\,[O/Fe] plot from Marino et al.\,(2015) is shown in panel (b).  Finally,  in panel (c) the red-RGB stars are colored either aqua or red for being considered the first or the second generation (1G,r and 2G,r)  of the iron-rich population, while panel (d) shows the corresponding  [Na/Fe] vs.\,[O/Fe] from Marino et al.\,(2015).}
 \label{fig:spettri2}
\end{figure}
\end{centering}
%%%%%%%%%%%%%%%%%%%%%%%%%%%%%%%%%%%%%%%%%%%%%%%%%%%%%%%%%%%%%%%%%%%%%%%%%%%%%%%

\section{Globular Clusters of Type II}
\label{sec:ano}

\begin{centering}
\begin{figure}
 \includegraphics[width=8.5cm]{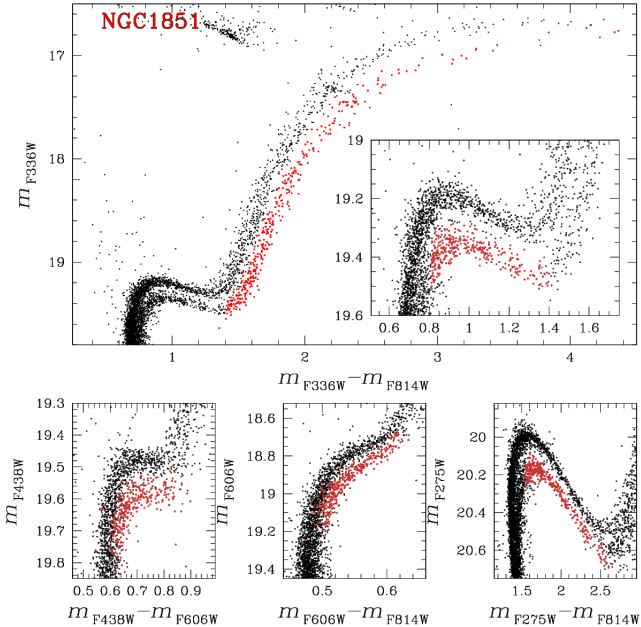}
 \caption{\textit{Upper panel:}  the  $m_{\rm F336W}$ vs.\, $m_{\rm F336W}-m_{\rm F814W}$ CMD of NGC\,1851 with red-RGB stars colored red. The inset shows a zoomed-in view around the SGB. \textit{Lower panels:} $m_{\rm F438W}$ vs.\,$m_{\rm F438W}-m_{\rm F814W}$ (left), $m_{\rm F606W}$ vs.\,$m_{\rm F606W}-m_{\rm F814W}$ (middle) and $m_{\rm F275W}$ vs.\,$m_{\rm F275W}-m_{\rm F814W}$ (right) CMDs around the SGB. The sample of faint SGB stars selected from the CMD in the insert of the upper-panel are colored red in these panels.}
 \label{fig:SGB1851}
\end{figure}
\end{centering}

In this section we present additional evidence to further explore and characterize the stellar-population content of Type II GCs, the most complex ones.
A visual inspection of the chromosome map of NGC\,1851 (Figure~\ref{fig:maps2}) reveals that the map itself appears to be split, with two 2G sequences running vaguely parallel to each other, and a hint of a second 1G sequence as well.  To better understand the origin of such a complex pattern, in 
Figure~\ref{fig:SGB1851} we show a collection of CMDs for NGC\,1851. The CMD in the  upper-panel   reveals that the SGB is clearly split into a bright and faint SGB (red points in the insert for the latter) which are connected to the blue and the red RGB, respectively. The RGB splitting was first noticed by Han et al.\,(2009)  using ground-based $U$ vs.\,$U-I$ photometry. 
 The red-RGB stars have been colored red in the upper panel of Figure~\ref{fig:SGB1851}.

We used the same colors to represent the sample of selected faint-SGB stars in the $m_{\rm F438W}$ vs.\, $m_{\rm F438W}-m_{\rm F814W}$, $m_{\rm F606W}$ vs.\,$m_{\rm F606W}-m_{\rm F814W}$ and $m_{\rm F275W}$ vs.\,$m_{\rm F275W}-m_{\rm F814W}$ CMDs,  plotted in the lower panels of Figure~\ref{fig:SGB1851}. These CMDs not only demonstrate  that the split SGB of NGC\,1851 is real, but also show that the faint SGB is visible also in CMDs made with optical filters, like $m_{\rm F606W}$ vs.\,$m_{\rm F606W}-m_{\rm F814W}$, where  stellar colors and luminosities are not significantly affected by  light-element abundance variations (see also Milone et al.\,2008). This indicates that faint-SGB stars are either enhanced in their C$+$N$+$O overall abundance, or are older than stars on the bright SGB by $\sim 1-2$Gyr (Cassisi et al.\,2008; Ventura et al.\,2009; Marino et al.\,2011). We emphasize that all type-II clusters exhibit either split or multimodal SGBs when observed in both ultraviolet and optical  filters, in contrast to Type I  GCs where multiple populations along the SGB are visible only in CMDs that include ultraviolet bands (Milone et al.\,2008; Marino et al.\,2009; Piotto et al.\,2012).

\begin{centering}
\begin{figure}
 \includegraphics[width=8.5cm]{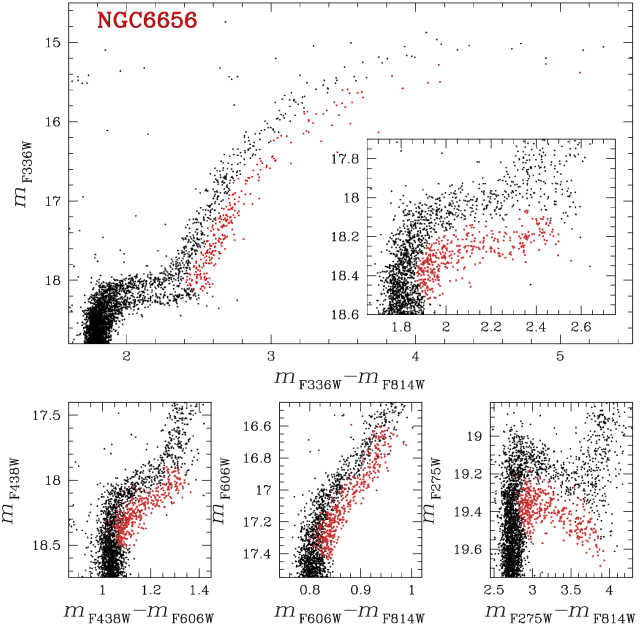}
 \caption{As in Figure~\ref{fig:SGB1851} but for NGC\,6656.}
 \label{fig:SGB6656}
\end{figure}
\end{centering}

\begin{centering}
\begin{figure}
 \includegraphics[width=8.5cm]{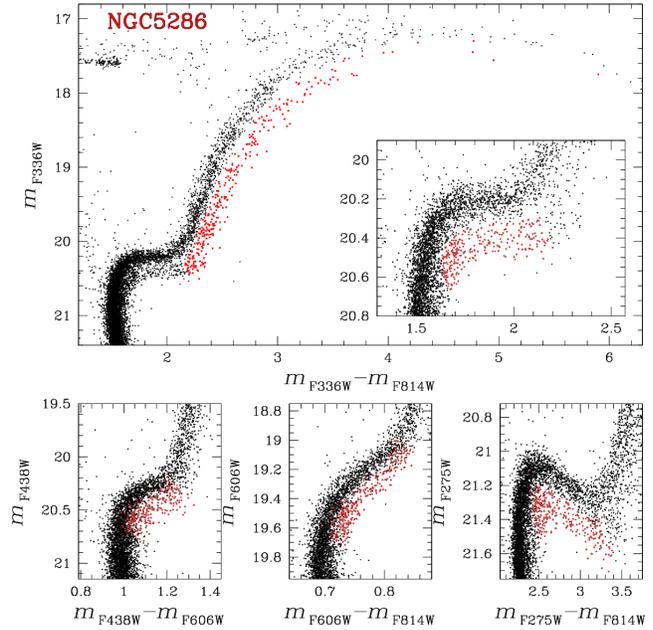}
 \caption{As in Figure~\ref{fig:SGB1851} but for NGC\,5286.}
 \label{fig:SGB5286}
\end{figure}
\end{centering}

\begin{centering}
\begin{figure}
 \includegraphics[width=8.5cm]{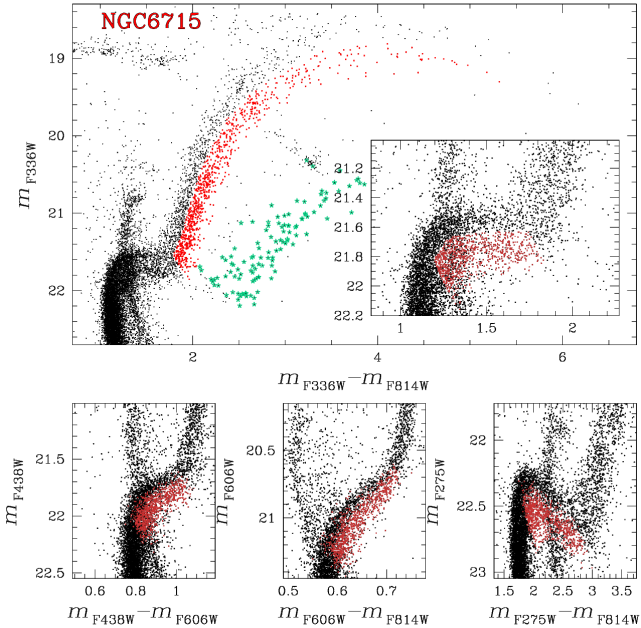}
 \caption{As in Figure~\ref{fig:SGB1851} but for NGC\,6715. This cluster (also known as M54) sits in the core of the Sagittarius dwarf galaxy, and the CMD in the upper panel shows stars of both the cluster and the core of this galaxy. In particular,  the extremely red RGB of the metal rich population of Sagittarius' core is recognizable just to the left of the inset (aqua starred symbols). These Sagittarius RGB stars are colored aqua in the chromosome map of NGC\,6715 shown in 
Figure~5.}
 \label{fig:SGB6715}
\end{figure}
\end{centering}

\begin{centering}
\begin{figure}
 \includegraphics[width=8.5cm]{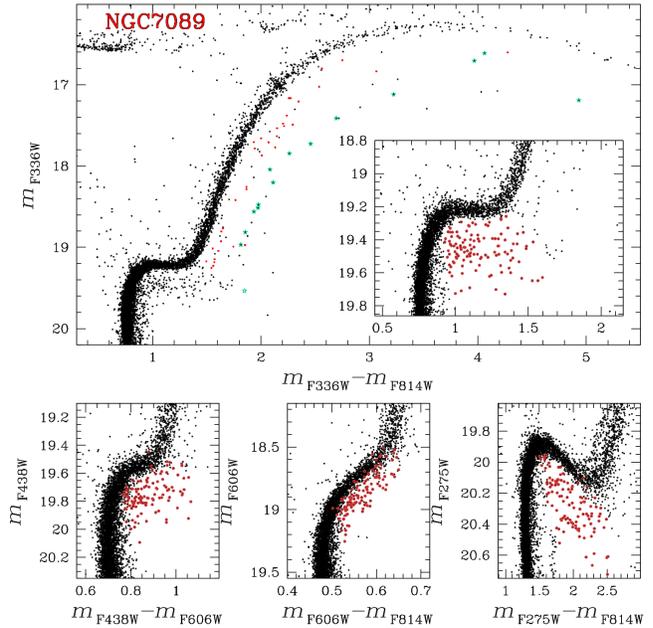}
 \caption{As in Figure~\ref{fig:SGB1851} but for NGC\,7089. The stars in the most-iron-rich RGB are represented with aqua stars.}
 \label{fig:SGB7089}
\end{figure}
\end{centering}

\begin{centering}
\begin{figure}
 \includegraphics[width=8.5cm]{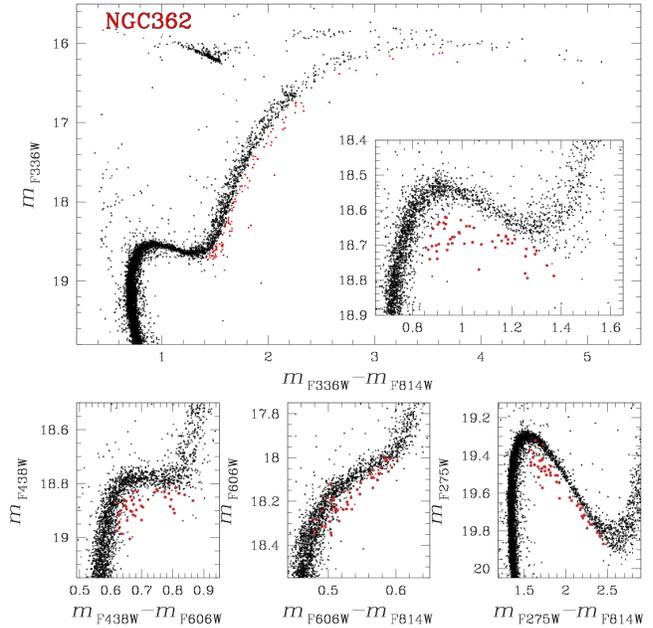}
 \caption{ As in Figure~\ref{fig:SGB1851} but for NGC\,362.}
 \label{fig:SGB0362}
\end{figure}
\end{centering}

\begin{centering}
\begin{figure}
 \includegraphics[width=8.5cm]{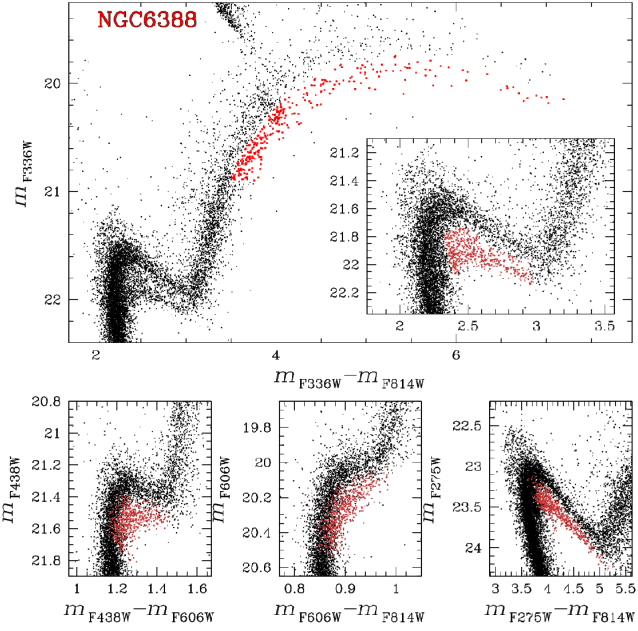}
 \caption{As in Figure~\ref{fig:SGB1851} but for NGC\,6388.}
 \label{fig:SGB6388}
\end{figure}
\end{centering}

\begin{centering}
\begin{figure}
 \includegraphics[width=8.5cm]{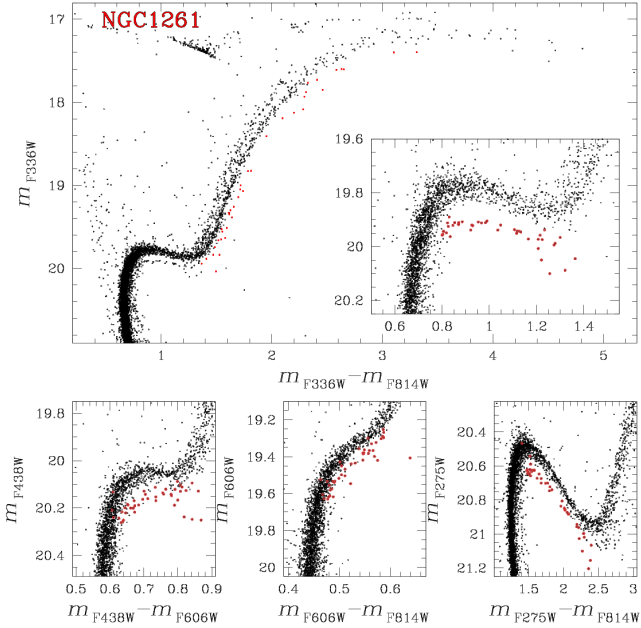}
 \caption{As in Figure~\ref{fig:SGB1851} but for NGC\,1261.}
 \label{fig:SGB1261}
\end{figure}
\end{centering}

\begin{centering}
\begin{figure}
 \includegraphics[width=8.5cm]{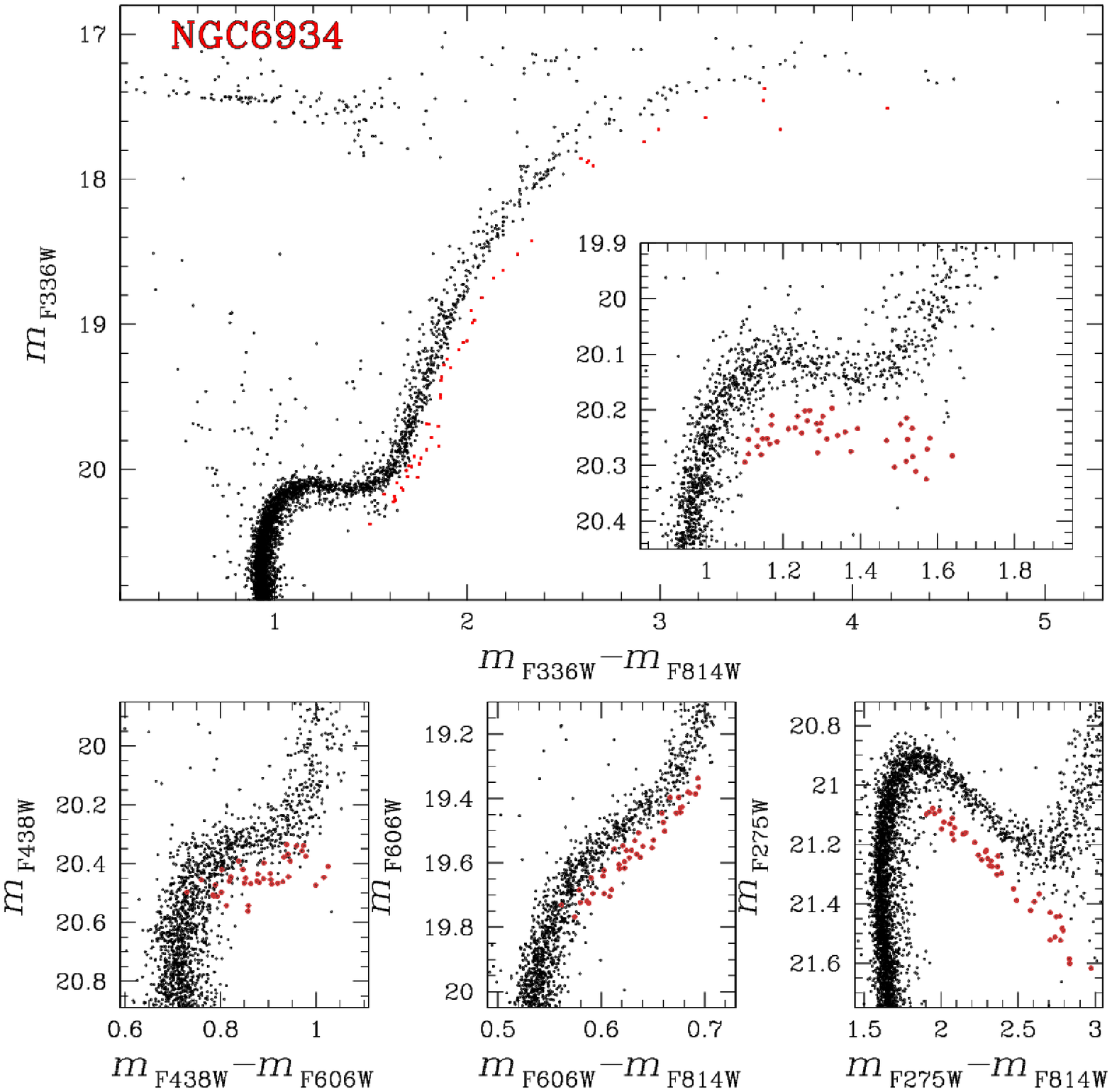}
 \caption{As in Figure~\ref{fig:SGB1851} but for NGC\,6934.}
 \label{fig:SGB6934}
\end{figure}
\end{centering}

A collection of CMDs for other Type II  clusters (namely, NGC\,362, NGC\,1261, NGC\,5139, NGC\,5286, NGC\,6656, NGC\,6715, NGC\,6093, and NGC\,7078) is provided in Figures~\ref{fig:SGB6656} to~\ref{fig:SGB6934}. Every CMD shows the existence a   faint SGB  that evolves into a red RGB in the $m_{\rm F336W}$ vs.\, $m_{\rm F336W}-m_{\rm F814W}$ CMD. As shown in Figure~\ref{fig:SGB6388}, the faint-SGB-RGB connection is unclear for NGC\,6388 where the RGB split is visible only for stars brighter than $m_{\rm F336W} \lesssim 20.75$. 
 Another possible exception  is 47\,Tucanae, in which there is no clear connection between multiple populations along the faint SGB and the RGB (Milone et al.\,2012b). The red-RGB stars identified in Figure~\ref{fig:SGB6656} to Figure~\ref{fig:SGB6934} are colored in red in the chromosome maps shown in  Figures~\ref{fig:maps1} to~\ref{fig:maps5}. The fact that red-RGB stars are clearly separated from the majority of cluster members in the chromosome maps demonstrates that the $\Delta_{\rm C~ F275W,F336W,F438W}$ vs.\,$\Delta_{\rm F275W,F814W}$ diagram is an efficient tool to identify GCs with multiple SGBs in the optical bands. 

The fraction of red-RGB stars with respect to the total number of analyzed RGB stars  differs significantly from one Type II cluster to another, and ranges from a minimum of $\sim$4\% for NGC\,1261  and NGC\,7089 to a maximum of $\sim$46\% and $\sim$64\% for NGC\,6715 and NGC\,5139 ($\omega$ Centauri), coming almost to dominate the cluster. Given its complexity,  the special case of $\omega$ Centauri requires a somewhat more elaborate procedure for the measurement of the RGB width and the construction of its chromosome map, which is illustrated in Appendix A.

For Type II  GCs, we have determined the RGB widths $W_{\rm  C~  F275W, F336W, F438W}$ and $W_{{\it m}_{\rm F275W}-{\it m}_{\rm F814W}}$ as described in Section~\ref{sec:mappe} for NGC\,6723,  but both by using only stars belonging to the blue RGB and by using  all the RGB stars. The latter quantities are called  $W^{*}_{\rm  C~  F275W, F336W, F438W}$ and $W^{*}_{{\it m}_{\rm F275W}-{\it m}_{\rm F814W}}$. Both $W$ and $W^*$  width values are reported in Table~\ref{tab:w}, with $W^*$ values given in a second row for each of the Type II clusters.

In order to illustrate the chemical tagging  of multiple populations in Type II  clusters, we use NGC\,5286 as a prototype. In panel (c) of Figure~\ref{fig:spettri}, 
red-RGB stars of this cluster are colored red whereas large black filled circles and red triangles are used for those stars for which  Marino et al.\,(2015) have measured their  content of iron and s-process-elements,  as shown in panel (d). Stars with low iron and barium belong to the 1G and 2G of the blue RGB, colored in black, while the stars enhanced in [Fe/H] and [Ba/Fe] populate the red RGB.

In panels (a) and (c) of Figure~\ref{fig:spettri2} we show separately the stars of the blue and the red RGBs of NGC\,5286, and compare the position of stars in the chromosome map and in the Na-O plot, in close analogy with what was previously done for NGC\,6121. We find that both RGBs host 1G stars with primordial oxygen and sodium abundance, and  2G stars enriched in sodium and depleted in oxygen, as shown in panels (b) and (d). In in Panel (c)  we indicate 1G and 2G stars of the red RGB as 1G,r and 2G,r, respectively. This finding is consistent we the conclusion by Marino et al.\,(2015) that both the group of barium-rich and barium-poor stars of NGC\,5286 exhibit their own Na-O anticorrelation. 
In Paper\,II, we have reached a similar conclusion for NGC\,7089,  using the abundances of light elements, s-process elements, and [Fe/H]  from Yong et al.\,(2014).
NGC\,7089 hosts a population of stars highly enhanced in iron with respect to the majority of cluster members. Stars in the extreme population of NGC\,7089 and the metal-rich stars in the core of the Sagittarius dwarf galaxy which are within the WFC3/UVIS images of  NGC\,6715 have been represented with aqua starred symbols in the corresponding chromosome maps. 

One intriguing discovery of the last decade is that a small but still increasing number of GCs  host two or more distinct groups of stars with different content of iron and s-process elements (Marino et al.\,2015;  Johnson et al.\,2015; Yong at al.\,2016 and references therein), while the majority of clusters have in general homogeneous abundances of s-process elements and metallicity. Moreover, the s-process rich stars are also iron rich and, in the cases of NGC\,6656, NGC\,1851, and $\omega$\,Centauri, these stars are also
enhanced in their overall C$+$N$+$O abundance (Yong et al.\,2009, 2014; Marino et al.\,2011, 2012, 2015; Villanova et al.\,2014).  

The chemical tagging of multiple populations is still quite fragmentary, especially for Type II clusters. However, all the available evidence indicates that 
stars in the faint SGB and red RGB are enhanced in global C$+$N$+$O content, in iron and in s-process elements. We conclude that Type II clusters differ from Type I ones  in three aspects: the SGB of type II GCs splits in optical bands, they host multiple 1G and/or 2G sequences in the chromosome maps and they show a wide composition range in heavy elements. Of course, these three characteristics ought to be physically connected to each other. To the best of our current understanding, each of these three properties, separately, is sufficient to identify as such a Type II cluster. We refer the reader to paper by Marino et al.\,(2015) and reference therein for further discussion on the chemical composition of Type II GCs.

%%%%%%%%%%%%%%%%%%%%%%%%%%%%%%%%%%%%% FIG 2 %%%%%%%%%%%%%%%%%%%%%%%%%%%%%%%%%%%
%%%%%%%%%%%%%%%%%%%%%%%%%%%%%%%%%%%%%%%%%%%%%%%%%%%%%%%%%%%%%%%%%%%%%%%%%%%%%%%

%%%%%%%%%%%%%%%%%%%%%%%%%%%%%%%%%%%%% FIG 2 %%%%%%%%%%%%%%%%%%%%%%%%%%%%%%%%%%%
%%%%%%%%%%%%%%%%%%%%%%%%%%%%%%%%%%%%%%%%%%%%%%%%%%%%%%%%%%%%%%%%%%%%%%%%%%%%%%%

%%%%%%%%%%%%%%%%%%%%%%%%%%%%%%%%%%%%% FIG 2 %%%%%%%%%%%%%%%%%%%%%%%%%%%%%%%%%%%
%%%%%%%%%%%%%%%%%%%%%%%%%%%%%%%%%%%%%%%%%%%%%%%%%%%%%%%%%%%%%%%%%%%%%%%%%%%%%%%

%%%%%%%%%%%%%%%%%%%%%%%%%%%%%%%%%%%%% FIG 2 %%%%%%%%%%%%%%%%%%%%%%%%%%%%%%%%%%%
%%%%%%%%%%%%%%%%%%%%%%%%%%%%%%%%%%%%%%%%%%%%%%%%%%%%%%%%%%%%%%%%%%%%%%%%%%%%%%%

%%%%%%%%%%%%%%%%%%%%%%%%%%%%%%%%%%%%% FIG 2 %%%%%%%%%%%%%%%%%%%%%%%%%%%%%%%%%%%
%%%%%%%%%%%%%%%%%%%%%%%%%%%%%%%%%%%%%%%%%%%%%%%%%%%%%%%%%%%%%%%%%%%%%%%%%%%%%%%

%%%%%%%%%%%%%%%%%%%%%%%%%%%%%%%%%%%%% FIG 2 %%%%%%%%%%%%%%%%%%%%%%%%%%%%%%%%%%%
%%%%%%%%%%%%%%%%%%%%%%%%%%%%%%%%%%%%%%%%%%%%%%%%%%%%%%%%%%%%%%%%%%%%%%%%%%%%%%%

%%%%%%%%%%%%%%%%%%%%%%%%%%%%%%%%%%%%% FIG 2 %%%%%%%%%%%%%%%%%%%%%%%%%%%%%%%%%%%
%%%%%%%%%%%%%%%%%%%%%%%%%%%%%%%%%%%%%%%%%%%%%%%%%%%%%%%%%%%%%%%%%%%%%%%%%%%%%%%

%%%%%%%%%%%%%%%%%%%%%%%%%%%%%%%%%%%%% FIG 2 %%%%%%%%%%%%%%%%%%%%%%%%%%%%%%%%%%%
%%%%%%%%%%%%%%%%%%%%%%%%%%%%%%%%%%%%%%%%%%%%%%%%%%%%%%%%%%%%%%%%%%%%%%%%%%%%%%%

%%%%%%%%%%%%%%%%%%%%%%%%%%%%%%%%%%%%% FIG 2 %%%%%%%%%%%%%%%%%%%%%%%%%%%%%%%%%%%
%%%%%%%%%%%%%%%%%%%%%%%%%%%%%%%%%%%%%%%%%%%%%%%%%%%%%%%%%%%%%%%%%%%%%%%%%%%%%%%

\section{Univariate relations between multiple populations and global cluster parameters} 
\label{sec:relW}
In this Section, we investigate the correlation between  the $W_{\rm C~ F275W, F336W, F438W}$ and $W_{\rm F275W, F814W}$ RGB widths and the 1G fraction, 
as determined in Section~\ref{sec:map}, and the global  parameters of the host GCs.  Such global GC parameters  include: 
metallicity ([Fe/H]), absolute visual magnitude ($M_{\rm V}$), central velocity dispersion ($\sigma_{\rm v}$), ellipticity ($\epsilon$), central concentration ($c$), core relaxation time ($\tau_{\rm c}$), half-mass relaxation time($\tau_{\rm hm}$), central stellar density ($\rho_{0}$), central surface brightness ($\mu_{\rm V}$), reddening ($E(B-V)$), and Galactocentric distance ($R_{\rm GC}$). All these quantities  are taken from the 2010 edition of the Harris\,(1996) catalog. 

The cluster masses have been taken from McLaughlin \& van der Marel\,(2005) for 44 of the GCs studied in this paper. The results of our paper are based on the masses obtained by fitting the models by Wilson\,(1975) on the profiles of 63 Galactic GCs by Trager et al.\,(1995). 
The fraction of binary stars in  GCs  has been taken from  Milone et al.\,(2012a), as measured within the cluster core ($f_{\rm bin}^{\rm C}$),  in the region between the core and the half-mass radius ($f_{\rm bin}^{\rm C-HM}$), and beyond  the half-mass radius ($f_{\rm bin}^{\rm oHM}$).

GC ages have been taken from Mar{\'i}n Franch et al.\,(2009, MF09), Dotter et al.\,(2010, D10) and Vandenberg et al.\,(2013, V13). All ages were obtained by using the same ACS/WFC dataset from Sarajedini et al.\,(2007) and Anderson et al.\,(2008) that we used in this paper, but different authors employed different sets of isochrones. 
 The Dotter et al.\,(2010) sample includes 50 of the GCs studied  in this paper. Additional ages for  six other GCs  were derived by Aaron Dotter by using the same method and are published in  Milone et al.\,(2014).

 The most recent age compilation comes from Vandenberg et al.\,(2013) and is based on an improved version of the classical `vertical method', i.e.\,the luminosity difference between the zero-age HB and the  main sequence turnoff. These authors have compared Victoria-Regina isochrones with GO-10775 photometry to derive  the ages for 51 of the GCs that we have analyzed in this  paper. 

When comparing two variables, we estimate the statistical correlation between the two by using the Spearman's rank correlation coefficient, $r$.  Moreover, we associate to each value of $r$ an uncertainty that is determined by  bootstrapping statistics as in Milone et al.\,(2014). Briefly, we have generated 1,000 equal-size resamples of the original dataset by randomly sampling with replacement from the observed dataset. For each $i$-th resample, we have determined $r_{\rm i}$ and considered the 68.27$^{\rm th}$ percentile of the $r_{\rm i}$ measurements ($\sigma_{\rm r}$) as indicative of the robustness of $r$.  

\subsection{RGB width and global  cluster parameters}
\label{sub:RelRGB}

\begin{centering}
\begin{figure}
 \includegraphics[width=8.7cm]{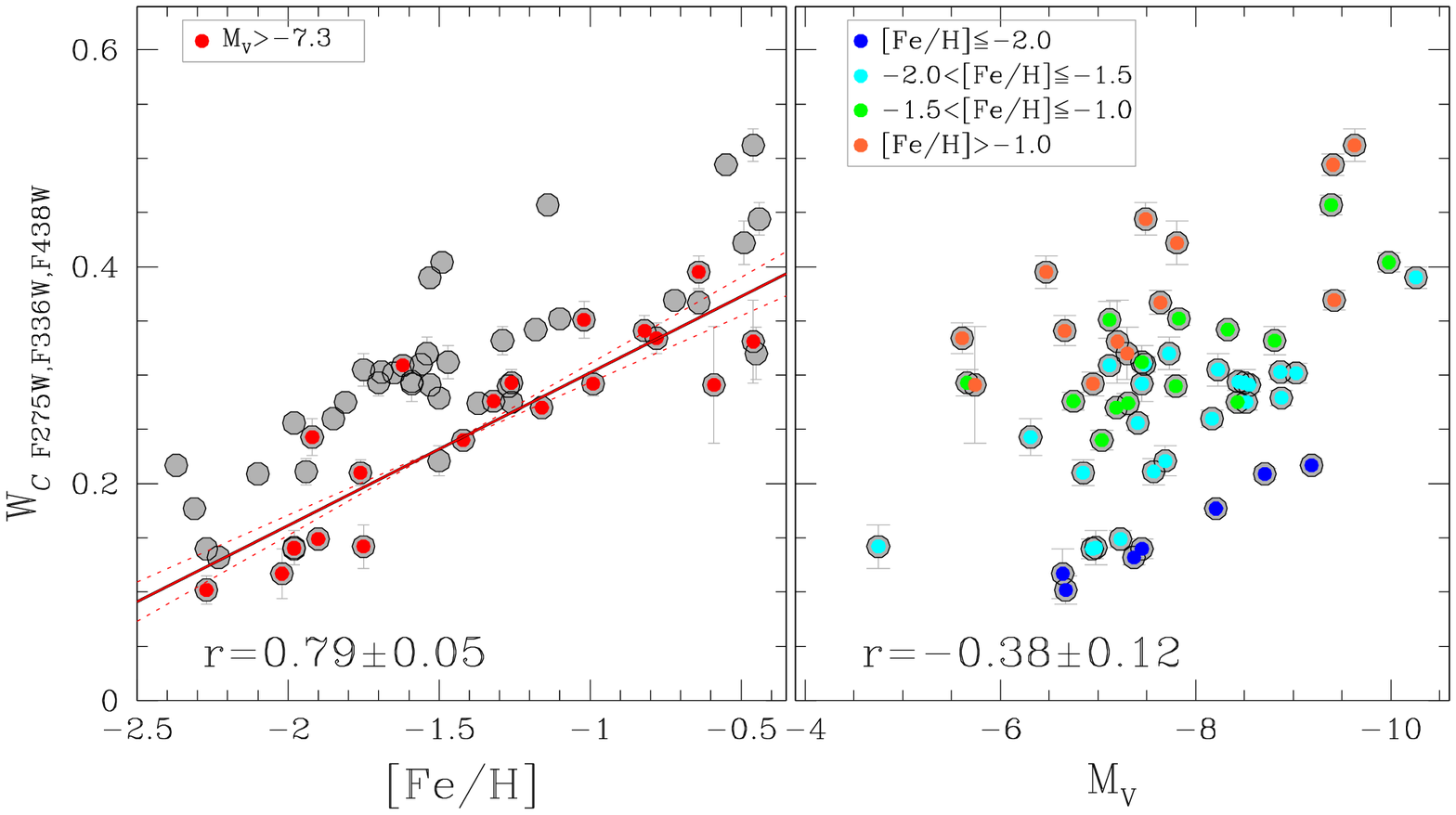}
 \includegraphics[width=8.7cm]{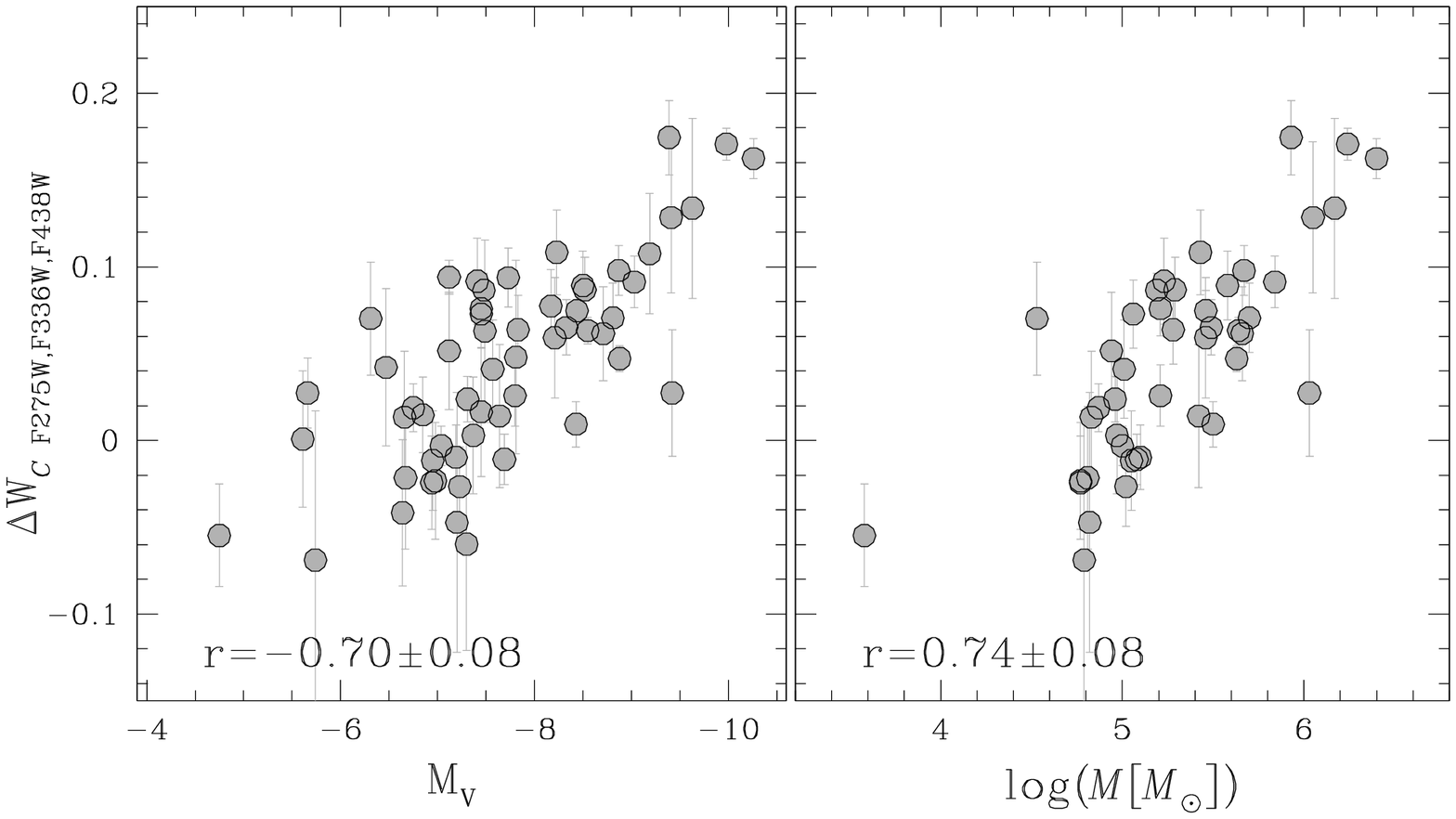}
%/home/milone/WORKS/treasury13/W/fig.macro go
%/home/milone/WORKS/treasury13/W/fig.macro go2
 \caption{{\it Upper-Left Panel:} the intrinsic RGB width, $W_{\rm C~ F275W, F336W, F438W}$, as a function of the iron abundance of the host GCs. The red line is the least-square  best-fit for the faint, less massive clusters with absolute magnitude $M_{\rm V}>-7.3$, that are marked with red dots. 
{\it Upper-Right Panel:}  the $W_{\rm C~ F275W, F336W, F438W}$ RGB width vs. the absolute visual magnitude $M_{\rm V}$ of the host clusters.  Clusters are color-coded depending on their metallicity [Fe/H] as indicated in the insert. 
{\it Lower Panels:}  the residuals of the RGB width, $\Delta W_{\rm C~ F275W, F336W, F438W}$, against the absolute magnitude (left) and the mass (right) of the host clusters. The Spearman's rank correlation coefficient ($r$) and the corresponding uncertainty are reported in each panel.}
 \label{fig:WfeMv}
\end{figure}
\end{centering}

Table~\ref{tab:correlazioni} lists the   Spearman's  rank correlation coefficients of the  $W_{\rm C~ F275W, F336W, F438W}$ RGB width with all the GC global parameters just listed above.  The Table also provides the number of clusters used for each determination of $r$, given in each column after the error on $r$.
 There is no significant correlation between the intrinsic RGB width and most of the global parameters, but a 
 strong correlation ($r=$0.79$\pm$0.05)  exists  between $W_{\rm C~ F275W, F336W, F438W}$ and metallicity, as shown in the left panel of Figure~\ref{fig:WfeMv}. This is hardly surprising, as at low metallicity the RGB colors  become almost insensitive to metal abundances while the RGB-color sensitivity to composition increases with increasing metallicity. 
 
There is only a mild correlation between the RGB width and the cluster absolute luminosity  ($r=-$0.38$\pm$0.12),  when using  the entire sample of GCs, as shown in the  right panel of Figure~\ref{fig:WfeMv}.
However, we note that GCs with almost the same [Fe/H] exhibit quite different $W_{\rm C~ F275W, F336W, F438W}$ values, thus suggesting that at least one more parameter is controlling the RGB width. Indeed, in the left panel of Figure~\ref{fig:WfeMv} we have marked with red dots GCs  fainter than $M_{\rm V}>-7.3$. Clearly the RGB width also depends on the cluster luminosity (or mass). 

 Low-mass clusters clearly exhibit, on average, smaller $W_{\rm C~ F275W, F336W, F438W}$ values than more-luminous, more-massive  GCs and define a tighter  $W_{\rm C~ F275W, F336W, F438W}$ vs.\,[Fe/H] correlation ($r=0.85\pm$0.07). 
 The significance of the correlation between RGB width and $M_{\rm V}$ becomes evident when distinguishing different metallicity ranges,
 as done in the right panel of Figure~\ref{fig:WfeMv}.   We found $r=0.76\pm$0.13 and $r=0.82\pm$0.11 for the selected groups of metal-rich and metal-poor GCs, respectively, and $r=0.73\pm$0.10 for GCs with $-2.0<$[Fe/H]$\leq -$1.5. The correlation coefficient has lower values for metal-intermediate GCs with $-1.5<$[Fe/H]$\leq -$1.0 and corresponds to $r=0.45\pm$0.22.

To further investigate the correlation between the width $W_{\rm C~ F275W, F336W, F438W}$ and global cluster parameters we 
need to remove the dependence on metallicity.  Thus, we have  least-squares fitted the $W_{\rm C~ F275W, F336W, F438W}$ vs.\, [Fe/H] relation for GCs with $M_{\rm V}>-7.3$ with a straight line,  as shown in the left panel of Figure~\ref{fig:WfeMv}, where the best-fit line is given by  $W_{\rm C~ F275W, F336W, F438W}=0.14\pm$0.02 [Fe/H]+0.44$\pm$0.03. We have then calculated  the residuals $\Delta W_{\rm C~ F275W, F336W, F438W}$ with respect to this line.
The values of the resulting Spearman's  rank correlation coefficient are listed in Table~\ref{tab:correlazioni} for each relation involving $\Delta W_{\rm C~ F275W, F336W, F438W}$. As expected, $\Delta W_{\rm C~ F275W, F336W, F438W}$ strongly correlates with the absolute luminosity and with the cluster mass (lower panels of Figure~\ref{fig:WfeMv} ($r > 0.7$).
 
%%%%%%%%%%%%%%%%%%%%%%%%%%%%%%%%%%%%% FIG 2 %%%%%%%%%%%%%%%%%%%%%%%%%%%%%%%%%%%
%%%%%%%%%%%%%%%%%%%%%%%%%%%%%%%%%%%%%%%%%%%%%%%%%%%%%%%%%%%%%%%%%%%%%%%%%%%%%%%

%%%%%%%%%%%%%%%%%%%%%%%%%%%%%%%%%%%%% FIG 2 %%%%%%%%%%%%%%%%%%%%%%%%%%%%%%%%%%%
\begin{centering}
\begin{figure*}
  \includegraphics[width=4.81cm]{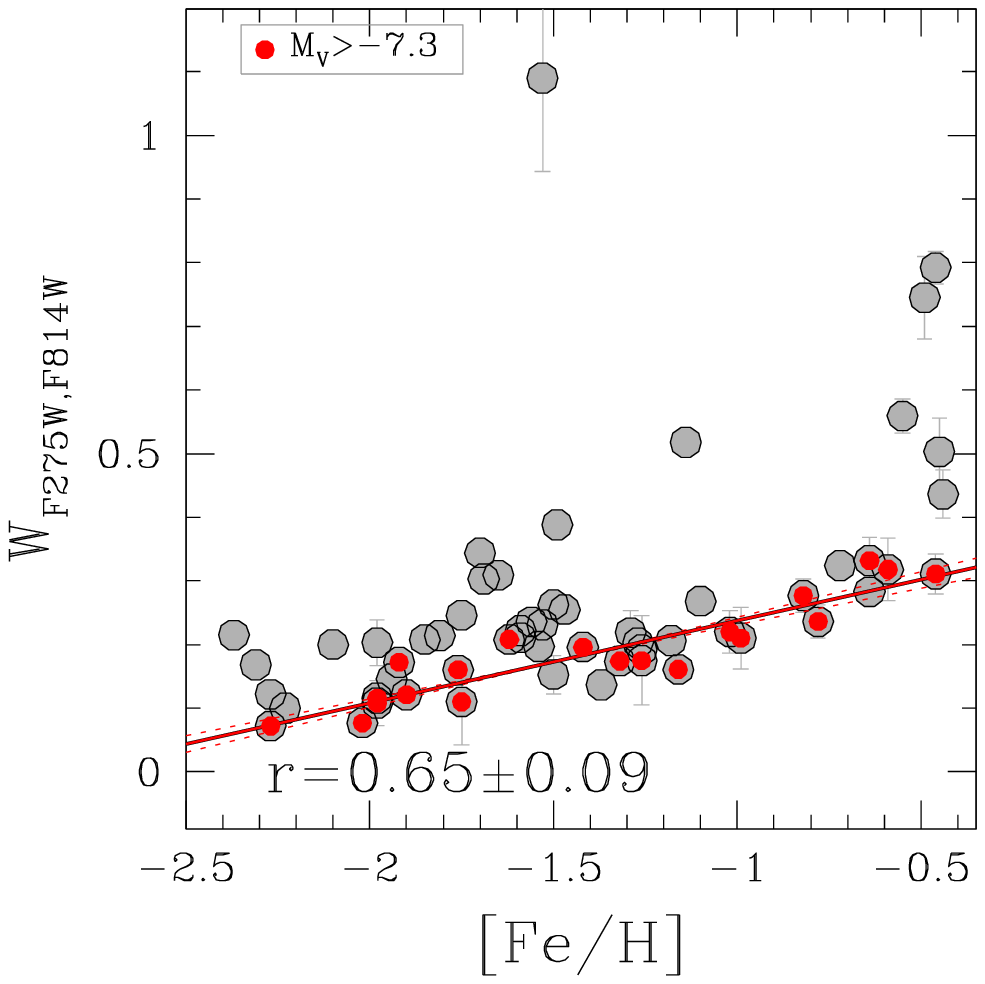}
  \includegraphics[width=8.7cm]{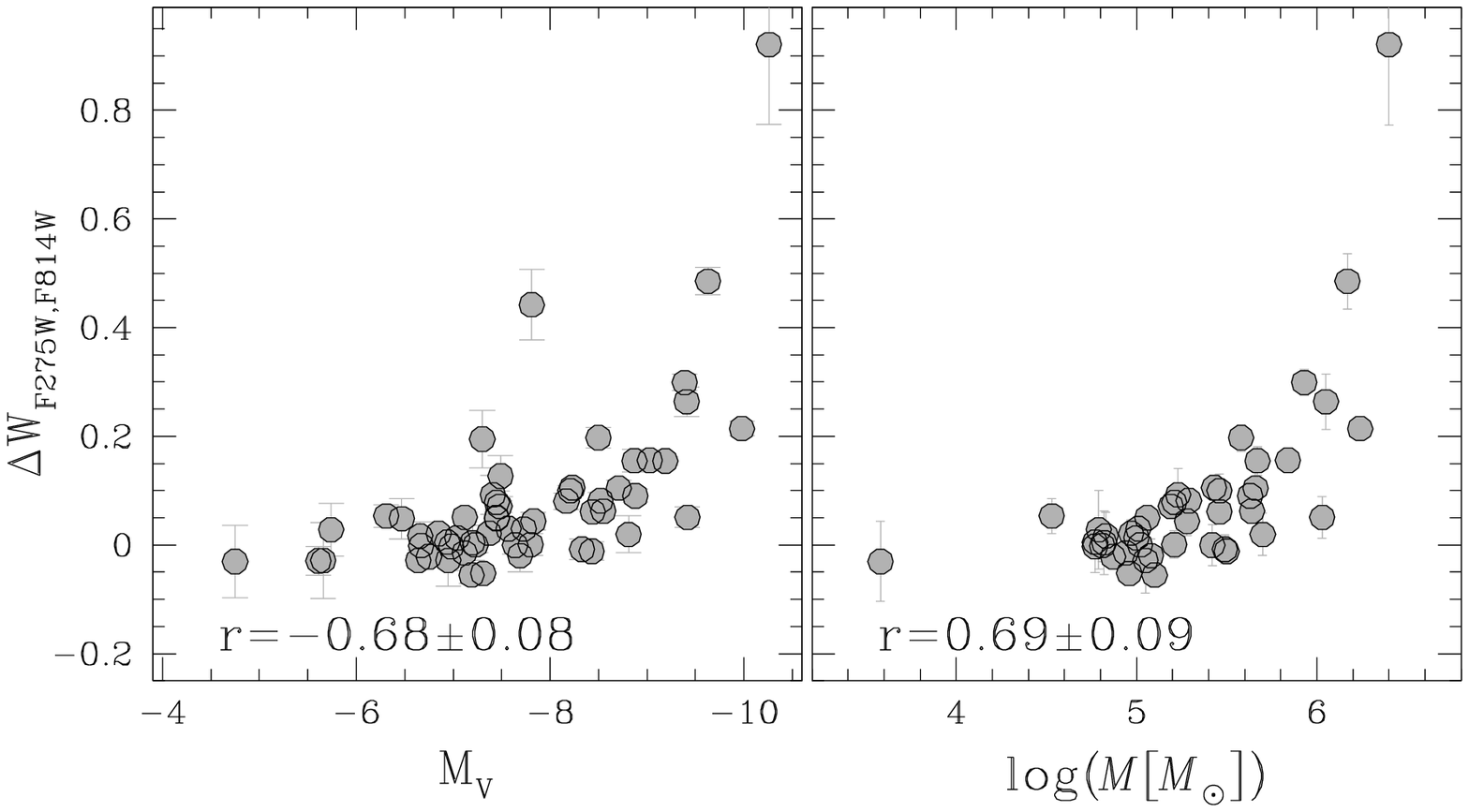}
%/home/milone/WORKS/treasury13/W/figW.macro go
\caption{The left panel shows the intrinsic RGB width, $W_{\rm F275W, F814W}$, as a  function of metallicity  of the host GC. The red line is the best-fit straight line for clusters with $M_{\rm V}>-7.3$ that we have represented with red dots. 
 The residuals of the RGB width with respect to the best-fit line, $\Delta W_{\rm F275W, F814W}$, are plotted against the absolute visual magnitude and the cluster mass in the middle and the right panels, respectively. The Spearman's rank correlation coefficient and the corresponding uncertainty are reported in each panel.
The outlier point refers to $\omega$ Centauri. }
 \label{fig:Wfi}
\end{figure*}
\end{centering}
%%%%%%%%%%%%%%%%%%%%%%%%%%%%%%%%%%%%%%%%%%%%%%%%%%%%%%%%%%%%%%%%%%%%%%%%%%%%%%%

%%%%%%%%%%%%%%%%%%%%%%%%%%%%%%%%%%%%% FIG 2 %%%%%%%%%%%%%%%%%%%%%%%%%%%%%%%%%%%
\begin{centering}
\begin{figure}
 \includegraphics[width=8.7cm]{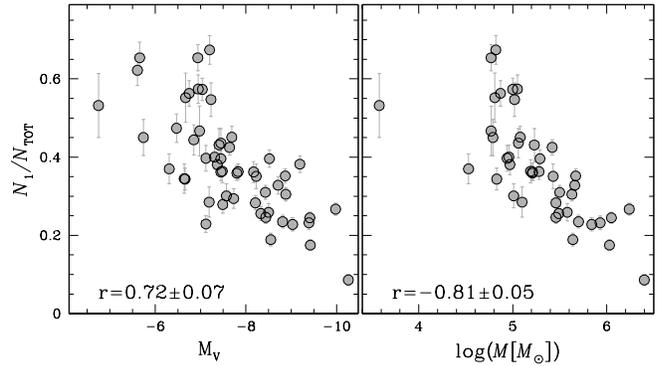}
%/home/milone/WORKS/treasury13/W/fig.macro go2p
%/home/milone/WORKS/treasury13/W/fig.macro go2p2
 \caption{ The fraction of 1G-stars with respect of the total number of used RGB stars as a function of the cluster absolute luminosity (left), cluster mass (right).} 
 \label{fig:pr12}
\end{figure}
\end{centering}
%%%%%%%%%%%%%%%%%%%%%%%%%%%%%%%%%%%%%%%%%%%%%%%%%%%%%%%%%%%%%%%%%%%%%%%%%%%%%%%

The $W_{\rm F275W, F814W}$ RGB width has then been analyzed in close analogy with what  done for $W_{\rm C~ F275W, F336W, F438W}$. As reported in Table~\ref{tab:correlazioni}, there is a positive correlation between $W_{\rm C~ F275W, F814W}$ and the cluster metallicity ($r=0.65\pm0.08$), see also 
the left panel of Figure~\ref{fig:Wfi}, where the less massive  clusters with $M_{\rm V}>-7.3$ are marked with red dots.  
The  least-squares best-fit straight line for the group of GCs with $M_{\rm V}>-7.3$ is plotted in red in the left panel of Figure~\ref{fig:Wfi} and  the residuals $\Delta W_{\rm F275W, F814W}$  with respect to such line are plotted as a function of cluster luminosity and mass in the two panels on the right of the same figure. 
Strong correlations of such residuals with cluster luminosity and mass are quite evident. 
 We have investigated the relation between $\Delta W_{\rm F275W, F814W}$ and the other global cluster parameters,  but no other significant correlation appears to exist, as reported in in Table~\ref{tab:correlazioni}.

\subsection{Fraction of 1G stars and  global cluster parameters}
\label{sec:relvarie}
 In this section we investigate univariate relations between the population ratio $N_1/N_{\rm TOT}$ and the global parameters of the host GCs, in analogy with what   done for the RGB width. The results are reported in Table~\ref{tab:correlazioni}. 

The most-relevant result is plotted in Figure~\ref{fig:pr12}, which shows significant anticorrelations between the  $N_1/N_{\rm TOT}$ ratio  and the absolute luminosity 
and  mass of the host clusters (with $r=-0.72\pm0.07$  and $r=-0.81\pm0.05$, respectively), with more-massive GCs having, on average, a smaller fraction of 1G stars. Based on a more limited data set it had been previously claimed that there is no correlation between the population ratio and cluster mass (Bastian \& Lardo 2015).
% Moreover, 
 On the contrary, the $N_1/N_{\rm TOT}$ ratio  correlates or anticorrelates with several quantities that are closely related with the cluster luminosity and mass.
The values of the Spearman's rank correlation coefficient listed in Table~\ref{tab:correlazioni} indicate a significant correlation with the central surface brightness 
($\mu_{\rm V}$, in magnitudes per square arc second, $r=0.71\pm0.07$) and significant anticorrelations with the central stellar density $\rho_0$ ($r=-0.63\pm0.09$) and the central velocity dispersion $\sigma_{\rm v}$ ($r=-0.63\pm0.09$).

We find no significant correlations between the fraction of 1G stars and other global parameters, in  particular between the population ratio and the distance from the Galactic center ($r=-0.05\pm0.13$) or with the cluster metallicity ($r=-0.08\pm0.15$). 

\subsection{The $\Delta_{\rm F275W, F814W}$ color extension of 1G and 2G stars and  global cluster parameters} 
We did not find any strong correlation between $W^{\rm 1G}_{\rm F275W, F814W}$ and any of the parameters that we have investigated. There is some mild correlation ($ r \sim 0.5$)  with the GC metallicity, the cluster mass and with GC ages ($r=-0.49 \pm 0.11$), but only when ages  from Vandenberg et al.\,(2013) are used.  
In summary,  it is  still unclear what controls the $\Delta_{\rm F275W, F814W}$ extension of 1G stars. 

\begin{centering}
\begin{figure}
 \includegraphics[width=8.7cm]{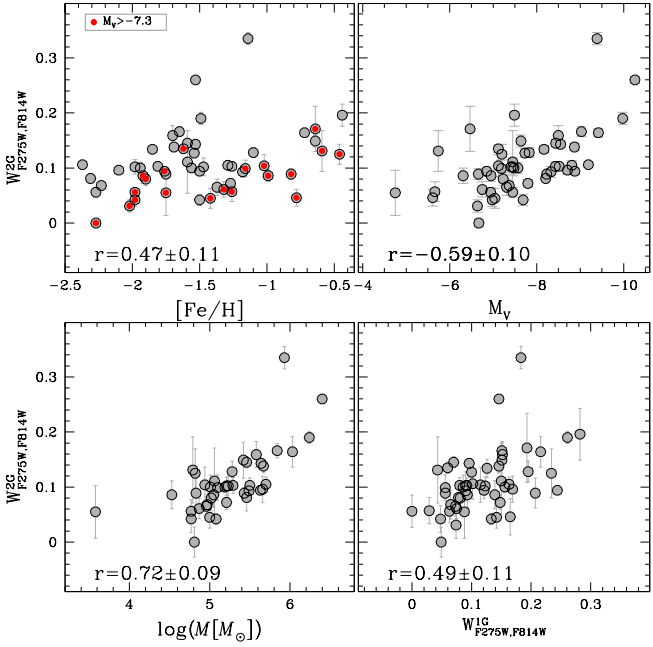}
%/home/milone/WORKS/treasury13/W/figWp.macro go
%/home/milone/WORKS/treasury13/NGC0288/MATCH/larg2pops.macro
\caption{The intrinsic width $W^{\rm 2G}_{\rm F275W, F814W}$  of the 2G stars as a function of cluster metallicity and absolute magnitude (upper left and right panels,
respectively),  and as a function of the cluster mass  and of the intrinsic width of 1G stars (lower left and right panels, respectively). Symbols are like in Figure~\ref{fig:WfeMv}. The Spearman's rank correlation coefficient and the corresponding uncertainty are indicated in each panel.}
 \label{fig:WpII}
\end{figure}
\end{centering}

In contrast, as shown in Figure~\ref{fig:WpII}, $W^{\rm 2G}_{\rm F275W, F814W}$  correlates with cluster mass and luminosity. Moreover, there is some anticorrelation with the fraction of 1G stars ($r=0.59 \pm 0.10$), which indicates that clusters with  a predominant 2G also have wide  RGB width in the F275W-F814W color. 
There is no strong correlation between $W^{\rm 2G}_{\rm F275W, F814W}$ and the cluster metallicity (r=0.35$\pm$0.11), although metal rich GCs with $M_{\rm V}>-7.3$ have on average larger values of $W^{\rm 2G}_{\rm F275W, F814W}$ than metal poor clusters within the same luminosity range. Similarly, there is only a mild  correlation between  the extension of the two generations,  $W^{\rm 2G}_{\rm F275W, F814W}$ and  $W^{\rm 1G}_{\rm F275W, F814W}$ ($r=0.49\pm 0.11$).

%%%%%%%%%%%%%%%%%%%%%%%%%%%%%%%%%%%%% FIG 2 %%%%%%%%%%%%%%%%%%%%%%%%%%%%%%%%%%%
%%%%%%%%%%%%%%%%%%%%%%%%%%%%%%%%%%%%%%%%%%%%%%%%%%%%%%%%%%%%%%%%%%%%%%%%%%%%%%%a

\section{Summary and conclusions}
\label{sec:summary}
We have analyzed high-precision multi-band {\it HST} photometry of 57 GCs in
order to identify and characterize their multiple stellar
populations along the RGB. The photometry has been collected through the F275W, F336W, F438W
filters of WFC3/UVIS and the F606W and F814W filters of WFC/ACS mostly as part of the {\it HST} UV Legacy Survey of Galactic GCs (Piotto et al.\,2015). Archive data have also been used. The main results can be summarized as follows.

\begin{itemize}
 \item From  the $m_{\rm F814W}$ vs.\,$C_{\rm F275W, F336W, F438W}$
 pseudo-CMD and  the $m_{\rm F814W}$ vs.\,$m_{\rm F275W}-m_{\rm F814W}$
 CMD of each cluster, which are both very sensitive to multiple stellar
 populations, we have  calculated the RGB width in $C_{\rm F275W, F336W, F438W}$ ($W_{\rm C~ F275W,F336W,F438W}$) and in $m_{\rm F275W}-m_{\rm F814W}$ ($W_{\rm F275W,F814W}$). In all 57 GCs, the
 observed RGB width is significantly wider than expected  from
 observational errors alone. This demonstrates that none of the studied
 GCs is consistent with hosting a simple stellar population.  Among them,
 $\omega$\,Centauri, ($\mathcal{M}=10^{6.13}\mathcal{M}_{\odot}$) and NGC\,6535 
 ($\mathcal{M}=10^{3.31} \mathcal{M}_{\odot}$) are respectively the most-massive and the  least-massive GC of the sample where multiple stellar populations have been detected to date.

 \item For each cluster we have combined the $C_{\rm F275W,F336W,F438W}$ pseudo-color and of the $m_{\rm F275W}-m_{\rm  F814W}$ color to construct  the $\Delta_{ \rm C~ F275W,F336W,F438W}$
 vs.\,$\Delta_{\rm F275W,F814W}$ pseudo two-color diagram,  or `chromosome map', 
 which maximizes the information on multiple stellar populations.
 
\item
The chromosome maps of the majority of the GCs shows two major, well separated 
  groups of stars, that we identify with first and second generation (1G and 2G). 1G stars are distributed around the
  origin of the chromosome map and span a narrow range of $\Delta_{\rm  C~ F275W,F336W,F438W}$  values.  The  group of 2G stars  that includes the remaining RGB stars span a wide range of both $\Delta_{\rm  C~ F275W,F336W,F438W}$ and $\Delta_{\rm F275W,F814W}$ values. Such a clean  1G/2G separation is not possible  for a few GCs (namely, NGC\,5927, NGC\,6304, and NGC\,6441), where the two sequences appear to be inextricably merged into a single sequence.
  Collectively, these clusters (with or without a clear 1G/2G separation) are called Type I clusters. 
\item
The chromosome maps of a second group of clusters, called Type II clusters, show a more complex pattern, with an apparent split of both 1G and 2G sequences.
A careful examination of multi-band CMDs of all these clusters reveals that their SGBs are split also in purely optical CMDs, while the SGB of Type I GCs splits only in CMDs based on ultraviolet filters.
By using spectroscopic data from the literature, we showed that Type II clusters host populations that are also enriched in overall CNO abundance (C+N+O) and heavy elements, such as iron and s-process elements. In particular, it is shown that the faint SGB corresponds to the stellar population
 enhanced in heavy elements (e.g.\,Marino et al.\,2011). 
We argue that 1)  split 1G and 2G sequences in the chromosome maps, 2) split SGBs and 3) non uniformity of the iron and s-elements abundances must be physically connected to each other.  This evidence indicates that chromosome maps are an efficient tool to identify GCs with internal variations of heavy elements. In this way, we have identified two new type-II GCs, namely NGC\,1261 and NGC\,6934.

 \item
 We use spectroscopic evidence from the literature to show that the photometrically-selected 1G and 2G stars  are oxygen-rich and sodium-poor
 and oxygen-poor and sodium-rich, respectively, supporting our identification of 1G and 2G stars with the first and second stellar generation, respectively.
 However, the number of stars with both accurate {\it HST} multi-band photometry and spectroscopic chemical analysis is still quite scanty. An extensive chemical tagging of multiple populations identified on the chromosome maps is now becoming a major requirement to further progress in the field of stellar populations in GCs. 
 
 \item
  Noticeably, the color width of both 1G and 2G stars in most GCs is significantly wider than what observational errors would suggest.  Such evidence
  demonstrates that in most GCs even the first (1G) stellar generation is not consistent with a simple, chemically homogeneous stellar population.
  Again, spectroscopic chemical tagging of 1G stars is needed to identify the origin of their wide range  of $\Delta_{\rm F275W,F814W}$ values.
 
  \item  
  We have investigated univariate relations between the RGB width in the $W_{\rm
 F275W,F814W}$  color and in the $W_{\rm C~ F275W,F336W,F438W}$ pseudo-color and the main global parameters of the host GCs.  The RGB width mostly correlates with cluster metallicity. After removing the dependence on metallicity, 
 significant correlations emerge  between the RGB width and   cluster mass and luminosity.
 These results indicate that massive GCs exhibit more pronounced  internal variations of helium and light elements  compared with low-mass GCs. 

\item
 For each cluster the F275W-F814W color width of 1G and
  2G stars ($W^{\rm 1G (2G)}_{\rm F275W,F814W}$)  have been measured. No
 significant correlation has been recovered between $W^{\rm 1G}_{\rm F275W,F814W}$ and any of
  the global cluster parameters. In contrast $W^{\rm 2G}_{\rm F275W,F814W}$
  correlates with the cluster mass. 
 
\item We have measured the fraction of 1G RGB stars with respect to
  the total number of RGB stars. The $N_{\rm 1G}/N_{\rm TOT}$ ratio ranges
  from $\sim$0.08 in the case of $\omega$\,Centauri to $\sim$0.67. There is a
  significant anticorrelation between the fraction of 1G stars and
  the mass of the host cluster, with massive GCs hosting a smaller fraction of
  1G stars. Hence, the multiple population phenomenon appears to systematically  increase in incidence and complexity with increasing cluster mass

\item 
In some cases distinct stellar clumps are clearly present along the sequence of 1G and/or 2G stars, while in other clusters we observe a smooth distribution without evident clumps. However, a large number of stars is needed to unambiguously identify distinct sub-populations along the main sequences in the chromosome maps, as done in Paper II and Paper III for NGC\,7089 and NGC\,2808, respectively.

\end{itemize}

\newpage
%%%%%%%%%%%%%%%%%%%%%%%%%%%%%%%%%%%%%%%%%%%%%%%%%%%%%%%%%%%%%%%%%%%%%%%%%%      
\begin{table*}
\caption{ Values of the RGB width and of the fraction of 1G stars with respect to the total number of analyzed stars. We also indicate type I and type II clusters and the fraction of type II stars with respect to the total number of analyzed stars. The last two columns provide the number of analyzed RGB stars and the ratio between the maximum radial distance from the cluster center of the analyzed stars ($R_{\rm max}$) and the cluster half-light radius ($R_{\rm hl}$). For the type-II clusters we provide in the second row the values of the RGB width obtained by excluding red-RGB stars ($W^{*}_{\rm  C~  F275W, F336W, F438W}$ and $W^{*}_{{\it m}_{\rm F275W}-{\it m}_{\rm F814W}}$).}
\scriptsize{
\begin{tabular}{lccccccccc}                                                
\hline
ID        & $W_{\rm C~ F275W, F336W, F438W}$ & $W_{{\it m}_{\rm F275W}-{\it m}_{\rm F814W}}$ & $W^{\rm 1G}_{{\it m}_{\rm F275W}-{\it m}_{\rm F814W}}$ & $W^{\rm 2G}_{{\it m}_{\rm F275W}-{\it m}_{\rm F814W}}$ & ${\rm N}_{1}/{\rm N}_{\rm TOT}$ & Type  & ${\rm N}_{\rm Type II}/{\rm N}_{\rm TOT}$ & N$_{\rm stars}$ & $R_{\rm max}/R_{\rm hl}$. \\
\hline
NGC\,0104 &  0.369$\pm$0.009 &    0.324$\pm$0.019 &  0.216$\pm$0.023 &  0.164$\pm$0.008 &   0.175$\pm$0.009  &     I    & 0               & 1853  & 0.56 \\
NGC\,0288 &  0.276$\pm$0.008 &    0.174$\pm$0.009 &  0.075$\pm$0.008 &  0.061$\pm$0.014 &   0.542$\pm$0.031  &     I    & 0               &  223  & 0.89 \\
NGC\,0362 &  0.275$\pm$0.005 &    0.192$\pm$0.017 &  0.092$\pm$0.012 &  0.103$\pm$0.008 &   0.279$\pm$0.015  &     II   & 0.075$\pm$0.009 &  840  & 2.01 \\
          &  0.271$\pm$0.007 &    0.187$\pm$0.013 &                  &                  &                    &          &                 &       &      \\
NGC\,1261 &  0.290$\pm$0.010 &    0.203$\pm$0.020 &  0.148$\pm$0.025 &  0.072$\pm$0.007 &   0.359$\pm$0.016  &     II   & 0.038$\pm$0.006 &  891  & 2.35 \\
          &  0.281$\pm$0.010 &    0.203$\pm$0.020 &                  &                  &                    &          &                 &       &      \\
NGC\,1851 &  0.342$\pm$0.005 &    0.206$\pm$0.019 &  0.090$\pm$0.010 &  0.093$\pm$0.010 &   0.264$\pm$0.015  &     II   & 0.030$\pm$0.014 & 1022  & 3.00 \\
          &  0.289$\pm$0.010 &    0.182$\pm$0.019 &                  &                  &                    &          &                 &       &      \\
NGC\,2298 &  0.243$\pm$0.017 &    0.172$\pm$0.021 &  0.139$\pm$0.026 &  0.086$\pm$0.014 &   0.370$\pm$0.037  &     I    & 0               &  156  & 1.61 \\
NGC\,2808 &  0.457$\pm$0.009 &    0.518$\pm$0.015 &  0.183$\pm$0.017 &  0.335$\pm$0.011 &   0.232$\pm$0.014  &     I    & 0               & 2682  & 2.32 \\
NGC\,3201 &  0.292$\pm$0.016 &    0.211$\pm$0.012 &  0.150$\pm$0.040 &  0.111$\pm$0.057 &   0.436$\pm$0.036  &     I    & 0               &  169  & 0.52 \\
NGC\,4590 &  0.132$\pm$0.007 &    0.100$\pm$0.005 &  0.065$\pm$0.008 &  0.068$\pm$0.007 &   0.381$\pm$0.024  &     I    & 0               &  330  & 1.13 \\
NGC\,4833 &  0.260$\pm$0.008 &    0.208$\pm$0.015 &  0.126$\pm$0.012 &  0.134$\pm$0.007 &   0.362$\pm$0.025  &     I    & 0               &  401  & 0.73 \\
NGC\,5024 &  0.209$\pm$0.005 &    0.200$\pm$0.014 &  0.169$\pm$0.016 &  0.096$\pm$0.008 &   0.328$\pm$0.020  &     I    & 0               & 1081  & 1.35 \\
NGC\,5053 &  0.102$\pm$0.013 &    0.072$\pm$0.009 &  0.049$\pm$0.012 &  0.000$\pm$0.007 &   0.544$\pm$0.062  &     I    & 0               & 56    & 0.53 \\
NGC\,5139 &  0.390$\pm$0.010 &    1.090$\pm$0.147 &  0.146$\pm$0.011 &  0.260$\pm$0.006 &   0.086$\pm$0.010  &     II   & 0.640$\pm$0.018 & 3084  & 0.50 \\ %0.393
          &  0.372$\pm$0.010 &    0.254$\pm$0.005 &                  &                  &                    &          &                 &       &      \\ %0.393
NGC\,5272 &  0.279$\pm$0.007 &    0.263$\pm$0.012 &  0.244$\pm$0.014 &  0.094$\pm$0.006 &   0.305$\pm$0.014  &     I    & 0               & 1177  & 0.83 \\
NGC\,5286 &  0.303$\pm$0.007 &    0.303$\pm$0.021 &  0.146$\pm$0.010 &  0.138$\pm$0.007 &   0.342$\pm$0.015  &     II   & 0.167$\pm$0.010 & 1521  & 2.25 \\
          &  0.292$\pm$0.013 &    0.249$\pm$0.014 &                  &                  &                    &          &                 &       &      \\
NGC\,5466 &  0.141$\pm$0.016 &    0.108$\pm$0.035 &  0.048$\pm$0.029 &  0.042$\pm$0.012 &   0.467$\pm$0.063  &     I    & 0               & 62    & 0.67 \\
NGC\,5897 &  0.149$\pm$0.008 &    0.121$\pm$0.014 &  0.081$\pm$0.019 &  0.080$\pm$0.012 &   0.547$\pm$0.042  &     I    & 0               & 194   & 0.79 \\
NGC\,5904 &  0.332$\pm$0.013 &    0.219$\pm$0.034 &  0.163$\pm$0.033 &  0.105$\pm$0.008 &   0.235$\pm$0.013  &     I    & 0               & 965   & 0.90 \\
NGC\,5927 &  0.422$\pm$0.020 &    0.745$\pm$0.065 &  0.631$\pm$0.066 &  0.304$\pm$0.037 &         ---        &     I    & 0               & 583   & 1.52 \\
NGC\,5986 &  0.294$\pm$0.008 &    0.222$\pm$0.007 &  0.070$\pm$0.006 &  0.145$\pm$0.007 &   0.246$\pm$0.012  &     I    & 0               & 895   & 1.81 \\
NGC\,6093 &  0.305$\pm$0.015 &    0.246$\pm$0.007 &  0.090$\pm$0.008 &  0.159$\pm$0.012 &   0.351$\pm$0.029  &     I    & 0               & 668   & 2.52 \\
NGC\,6101 &  0.140$\pm$0.009 &    0.116$\pm$0.012 &  0.063$\pm$0.013 &  0.056$\pm$0.008 &   0.654$\pm$0.032  &     I    & 0               & 263   & 1.48 \\
NGC\,6121 &  0.270$\pm$0.007 &    0.161$\pm$0.015 &  0.056$\pm$0.045 &  0.099$\pm$0.015 &   0.285$\pm$0.037  &     I    & 0               & 135   & 0.39 \\
NGC\,6144 &  0.210$\pm$0.012 &    0.160$\pm$0.012 &  0.121$\pm$0.023 &  0.094$\pm$0.008 &   0.444$\pm$0.037  &     I    & 0               & 159   & 0.95 \\
NGC\,6171 &  0.351$\pm$0.017 &    0.220$\pm$0.033 &  0.115$\pm$0.020 &  0.104$\pm$0.020 &   0.397$\pm$0.031  &     I    & 0               & 245   & 0.90 \\
NGC\,6205 &  0.291$\pm$0.006 &    0.231$\pm$0.008 &  0.096$\pm$0.020 &  0.143$\pm$0.006 &   0.184$\pm$0.013  &     I    & 0               & 1198  & 1.05 \\
NGC\,6218 &  0.274$\pm$0.009 &    0.137$\pm$0.009 &  0.073$\pm$0.018 &  0.065$\pm$0.015 &   0.400$\pm$0.029  &     I    & 0               & 315   & 0.93 \\
NGC\,6254 &  0.310$\pm$0.007 &    0.236$\pm$0.011 &  0.156$\pm$0.020 &  0.100$\pm$0.008 &   0.364$\pm$0.028  &     I    & 0               & 488   & 0.86 \\
NGC\,6304 &  0.320$\pm$0.024 &    0.503$\pm$0.053 &  0.371$\pm$0.083 &  0.228$\pm$0.028 &         ---        &     I    & 0               & 602   & 1.13 \\
NGC\,6341 &  0.177$\pm$0.005 &    0.168$\pm$0.009 &  0.078$\pm$0.011 &  0.081$\pm$0.003 &   0.304$\pm$0.015  &     I    & 0               & 795   & 1.63 \\
NGC\,6352 &  0.395$\pm$0.015 &    0.332$\pm$0.037 &  0.193$\pm$0.053 &  0.171$\pm$0.041 &   0.474$\pm$0.035  &     I    & 0               & 221   & 0.76 \\
NGC\,6362 &  0.292$\pm$0.011 &    0.210$\pm$0.048 &  0.093$\pm$0.036 &  0.086$\pm$0.010 &   0.574$\pm$0.035  &     I    & 0               & 233   & 0.81 \\
NGC\,6366 &  0.291$\pm$0.064 &    0.318$\pm$0.049 &  0.043$\pm$0.075 &  0.131$\pm$0.037 &   0.418$\pm$0.045  &     I    & 0               & 72    & 0.51 \\
NGC\,6388 &  0.494$\pm$0.010 &    0.559$\pm$0.027 &       ---        &       ---        &   0.245$\pm$0.010  &     II   & 0.299$\pm$0.016 & 1735  & 2.45 \\ 
NGC\,6397 &  0.117$\pm$0.023 &    0.077$\pm$0.009 &  0.074$\pm$0.011 &  0.031$\pm$0.011 &   0.345$\pm$0.036  &     I    & 0               & 111   & 0.55 \\
NGC\,6441 &  0.512$\pm$0.015 &    0.792$\pm$0.025 &  0.283$\pm$0.025 &  0.298$\pm$0.017 &         ---        &     I    & 0               & 1907  & 2.90 \\
NGC\,6496 &  0.331$\pm$0.038 &    0.311$\pm$0.032 &  0.234$\pm$0.033 &  0.125$\pm$0.018 &   0.674$\pm$0.035  &     I    & 0               & 196   & 1.40 \\
NGC\,6535 &  0.142$\pm$0.020 &    0.110$\pm$0.067 &  0.088$\pm$0.015 &  0.055$\pm$0.041 &   0.536$\pm$0.081  &     I    & 0               & 62    & 1.70 \\
NGC\,6541 &  0.275$\pm$0.007 &    0.214$\pm$0.015 &  0.080$\pm$0.009 &  0.103$\pm$0.006 &   0.396$\pm$0.020  &     I    & 0               & 692   & 1.56 \\
NGC\,6584 &  0.221$\pm$0.014 &    0.153$\pm$0.030 &  0.133$\pm$0.031 &  0.042$\pm$0.010 &   0.451$\pm$0.026  &     I    & 0               & 417   & 2.27 \\
NGC\,6624 &  0.444$\pm$0.015 &    0.436$\pm$0.038 &  0.282$\pm$0.040 &  0.196$\pm$0.020 &   0.279$\pm$0.020  &     I    & 0               & 594   & 1.87 \\
NGC\,6637 &  0.367$\pm$0.011 &    0.283$\pm$0.016 &  0.151$\pm$0.022 &  0.149$\pm$0.011 &   0.425$\pm$0.017  &     I    & 0               & 862   & 2.05 \\
NGC\,6652 &  0.341$\pm$0.014 &    0.277$\pm$0.026 &  0.207$\pm$0.027 &  0.089$\pm$0.010 &   0.344$\pm$0.026  &     I    & 0               & 340   & 3.09 \\
NGC\,6656 &  0.293$\pm$0.012 &    0.344$\pm$0.019 &  0.152$\pm$0.030 &  0.159$\pm$0.018 &   0.274$\pm$0.020  &     II   & 0.403$\pm$0.021 & 557   & 0.51 \\
          &  0.215$\pm$0.010 &    0.234$\pm$0.023 &                  &                  &                    &          &                 &       &      \\
NGC\,6681 &  0.309$\pm$0.005 &    0.208$\pm$0.009 &  0.060$\pm$0.013 &  0.135$\pm$0.007 &   0.234$\pm$0.019  &     I    & 0               & 527   & 2.31 \\
NGC\,6715 &  0.404$\pm$0.009 &    0.388$\pm$0.013 &  0.261$\pm$0.016 &  0.190$\pm$0.011 &   0.267$\pm$0.012  &     II   & 0.046$\pm$0.011 & 2358  & 2.08 \\
          &  0.346$\pm$0.012 &    0.349$\pm$0.016 &                  &                  &                    &          &                 &       &      \\
NGC\,6717 &  0.293$\pm$0.012 &    0.175$\pm$0.070 &  0.029$\pm$0.015 &  0.057$\pm$0.018 &   0.637$\pm$0.039  &     I    & 0               & 102   & 2.01 \\   
NGC\,6723 &  0.352$\pm$0.006 &    0.268$\pm$0.016 &  0.195$\pm$0.020 &  0.128$\pm$0.009 &   0.363$\pm$0.017  &     I    & 0               & 695   & 1.05 \\
NGC\,6752 &  0.320$\pm$0.015 &    0.197$\pm$0.010 &  0.100$\pm$0.016 &  0.127$\pm$0.008 &   0.294$\pm$0.023  &     I    & 0               & 372   & 0.91 \\
NGC\,6779 &  0.256$\pm$0.007 &    0.203$\pm$0.036 &  0.090$\pm$0.039 &  0.102$\pm$0.013 &   0.469$\pm$0.041  &     I    & 0               & 420   & 1.29 \\
NGC\,6809 &  0.211$\pm$0.012 &    0.146$\pm$0.006 &  0.086$\pm$0.008 &  0.100$\pm$0.010 &   0.311$\pm$0.029  &     I    & 0               & 171   & 0.55 \\
NGC\,6838 &  0.334$\pm$0.014 &    0.236$\pm$0.026 &  0.165$\pm$0.025 &  0.046$\pm$0.015 &   0.622$\pm$0.038  &     I    & 0               & 132   & 0.88 \\
NGC\,6934 &  0.312$\pm$0.015 &    0.255$\pm$0.021 &  0.123$\pm$0.028 &  0.102$\pm$0.016 &   0.326$\pm$0.020  &     II   & 0.067$\pm$0.010 & 606   & 2.30 \\
          &  0.304$\pm$0.013 &    0.237$\pm$0.015 &                  &                  &                    &          &                 &       &      \\
NGC\,6981 &  0.240$\pm$0.009 &    0.196$\pm$0.019 &  0.142$\pm$0.026 &  0.045$\pm$0.018 &   0.542$\pm$0.027  &     I    & 0               & 389   & 1.67 \\
NGC\,7078 &  0.217$\pm$0.003 &    0.215$\pm$0.007 &  0.102$\pm$0.007 &  0.106$\pm$0.005 &   0.399$\pm$0.019  &     I    & 0               & 1495  & 1.79 \\
NGC\,7089 &  0.302$\pm$0.009 &    0.309$\pm$0.014 &  0.151$\pm$0.022 &  0.166$\pm$0.009 &   0.224$\pm$0.014  &     II   & 0.043$\pm$0.006 & 1296  & 1.47 \\
          &  0.302$\pm$0.009 &    0.309$\pm$0.014 &                  &                  &                    &          &                 &       &      \\
NGC\,7099 &  0.140$\pm$0.009 &    0.122$\pm$0.017 &  0.000$\pm$0.010 &  0.056$\pm$0.009 &   0.380$\pm$0.028  &     I    & 0               & 323   & 1.55 \\
\hline\hline
\end{tabular}
}
\label{tab:w}
\end{table*}
%%%%%%%%%%%%%%%%%%%%%%%%%%%%%%%%%%%%%%%%%%%%%%%%%%%%%%%%%%%%%%%%%%%%%%%%%%      

\begin{table*}
\caption{For each couple of quantities we list the Spearman's rank correlation coefficient, the corresponding uncertainty, and the number of clusters  used to calculate the correlation coefficient.}
\scriptsize{
\begin{tabular}{ccccccc}
\hline
 Parameter & $W_{C \rm F275W, F336W, F438W}$ & $\Delta W_{C \rm F275W, F336W, F438W}$ & $W^{*}_{C \rm F275W, F336W, F438W}$ & $\Delta W^{*}_{C \rm F275W, F336W, F438W}$ & $W_{\rm F275W, F814W}$ & $\Delta W_{\rm F275W, F814W}$    \\
\hline
 $\sigma_{\rm v}$            &      0.30$\pm$0.14, 57 &   0.63$\pm$0.08, 57  &   0.20$\pm$0.15, 56 &   0.54$\pm$0.10, 56 &    0.35$\pm$0.14, 57 &   0.46$\pm$0.12, 57    \\
 $c$                         &    0.23$\pm$0.14, 57 &   0.38$\pm$0.13, 57  &   0.18$\pm$0.14, 56 &   0.36$\pm$0.13, 56 &    0.17$\pm$0.14, 57 &   0.31$\pm$0.12, 57    \\
 $\mu_{\rm V}$               &   $-$0.44$\pm$0.12, 57 &$-$0.61$\pm$0.09, 57  &$-$0.37$\pm$0.12, 56 &$-$0.56$\pm$0.11, 56 & $-$0.41$\pm$0.13, 57 &$-$0.48$\pm$0.10, 57    \\
 $\epsilon$                  &    0.08$\pm$0.12, 57 &$-$0.02$\pm$0.14, 57  &$-$0.07$\pm$0.13, 56 &$-$0.04$\pm$0.14, 56 &    0.08$\pm$0.14, 57 &   0.15$\pm$0.14, 57    \\
 $\rho_{0}$                  &     0.44$\pm$0.12, 57 &   0.51$\pm$0.12, 57  &   0.37$\pm$0.14, 56 &   0.45$\pm$0.12, 56 &    0.39$\pm$0.13, 57 &   0.41$\pm$0.12, 57    \\
 $log{\tau_{\rm c}}$         &   $-$0.22$\pm$0.15, 57 &$-$0.14$\pm$0.15, 57  &$-$0.19$\pm$0.14, 56 &$-$0.12$\pm$0.15, 56 & $-$0.07$\pm$0.15, 57 &   0.01$\pm$0.14, 57    \\
 $log{\tau_{\rm hm}}$        &   $-$0.18$\pm$0.15, 57 &   0.17$\pm$0.13, 57  &$-$0.17$\pm$0.15, 56 &   0.19$\pm$0.14, 56 & $-$0.01$\pm$0.15, 57 &   0.26$\pm$0.13, 57    \\
 R$_{\rm GC}$                &   $-$0.38$\pm$0.12, 57 &   0.02$\pm$0.13, 57  &$-$0.41$\pm$0.12, 56 &   0.01$\pm$0.13, 56 & $-$0.30$\pm$0.11, 57 &   0.01$\pm$0.13, 57     \\
 age (MF09)                  &  $-$0.31$\pm$0.12, 56 &   0.07$\pm$0.13, 56  &$-$0.26$\pm$0.12, 55 &   0.11$\pm$0.13, 55 & $-$0.26$\pm$0.12, 56 &$-$0.05$\pm$0.14, 56    \\
 age (D10)                   &  $-$0.41$\pm$0.11, 56 &   0.06$\pm$0.15, 56  &$-$0.39$\pm$0.11, 55 &   0.11$\pm$0.14, 55 & $-$0.29$\pm$0.13, 56 &   0.11$\pm$0.15, 56    \\
 age (V13)                   &  $-$0.54$\pm$0.09, 51 &   0.17$\pm$0.14, 51  &$-$0.53$\pm$0.10, 51 &   0.18$\pm$0.15, 51 & $-$0.51$\pm$0.10, 51 &   0.01$\pm$0.15, 51     \\
  {\rm [Fe/H]}               &     0.79$\pm$0.05, 57 &$-$0.07$\pm$0.14, 57  &   0.79$\pm$0.05, 56 &$-$0.11$\pm$0.14, 56 &    0.65$\pm$0.09, 57 &$-$0.03$\pm$0.14, 57     \\
 M$_{\rm V}$                 &   $-$0.38$\pm$0.12, 57 &$-$0.70$\pm$0.07, 57  &$-$0.29$\pm$0.14, 56 &$-$0.64$\pm$0.09, 56 & $-$0.50$\pm$0.12, 57 &$-$0.68$\pm$0.08, 57    \\
 log{$\mathcal{M/M_{\odot}}$}&      0.60$\pm$0.12, 44 &   0.74$\pm$0.08, 44  &   0.51$\pm$0.13, 43 &   0.68$\pm$0.10, 43 &    0.65$\pm$0.12, 44 &   0.69$\pm$0.09, 44   \\
 $f^{\rm C}_{\rm bin}$        &    0.18$\pm$0.17, 34 &$-$0.40$\pm$0.15, 34  &   0.23$\pm$0.17, 34 &$-$0.32$\pm$0.15, 34 &    0.12$\pm$0.18, 34 &$-$0.36$\pm$0.14, 34       \\
 $f^{\rm C-HM}_{\rm bin}$     & $-$0.08$\pm$0.15, 46 &$-$0.44$\pm$0.12, 46  &$-$0.06$\pm$0.15, 46 &$-$0.42$\pm$0.13, 46 & $-$0.13$\pm$0.16, 46 &$-$0.42$\pm$0.11, 46         \\
 $f^{\rm oHM}_{\rm bin}$     &  $-$0.29$\pm$0.16, 42 &$-$0.51$\pm$0.13, 42  &$-$0.22$\pm$0.16, 41 &$-$0.44$\pm$0.13, 41 & $-$0.26$\pm$0.15, 42 &$-$0.37$\pm$0.14, 42         \\
 $S_{\rm RR~Lyrae}$         &   $-$0.24$\pm$0.12, 57 &   0.02$\pm$0.15, 57  &$-$0.26$\pm$0.13, 56 &$-$0.01$\pm$0.15, 56 & $-$0.23$\pm$0.13, 57 &$-$0.14$\pm$0.14, 57          \\
 $E(B-V)$                &       0.34$\pm$0.12, 57 &$-$0.11$\pm$0.14, 57  &   0.31$\pm$0.12, 56 &   0.06$\pm$0.14, 56 &    0.41$\pm$0.12, 57 &   0.22$\pm$0.14, 57       \\
 $N_{\rm 1}/N_{\rm TOT}$    &   $-$0.41$\pm$0.12, 54 &$-$0.61$\pm$0.09, 54  &$-$0.32$\pm$0.13, 53 &$-$0.54$\pm$0.10, 53 & $-$0.44$\pm$0.12, 54 &$-$0.56$\pm$0.09, 54       \\
\hline
\hline
Parameter &  $W^{*}_{\rm F275W, F814W}$ & $\Delta W^{*}_{\rm F275W, F814W}$ &   $N_{\rm 1}/N_{\rm TOT}$  & $W^{\rm 1G}_{\rm F275W, F814W}$  & $W^{\rm 2G}_{\rm F275W, F814W}$ &  \\
\hline
$\sigma_{\rm V}$     &    0.26$\pm$0.14, 56 &    0.39$\pm$0.13, 56  & $-$0.63$\pm$0.09, 54 &   0.12$\pm$0.15, 53 &   0.39$\pm$0.13, 53 &                \\
$c$                 &    0.17$\pm$0.14, 56 &    0.32$\pm$0.12, 56  & $-$0.54$\pm$0.11, 54 &   0.08$\pm$0.17, 53 &   0.18$\pm$0.14, 53 &               \\
$\mu_{\rm V}$         & $-$0.38$\pm$0.13, 56 & $-$0.46$\pm$0.11, 56  &    0.71$\pm$0.07, 54 &$-$0.19$\pm$0.14, 53 &$-$0.42$\pm$0.12, 53 &                \\
$\epsilon$          &    0.07$\pm$0.13, 56 &    0.13$\pm$0.13, 56  & $-$0.07$\pm$0.14, 54 &   0.15$\pm$0.14, 53 &   0.24$\pm$0.13, 53 &               \\
$\rho_{0}$          &    0.37$\pm$0.14, 56 &    0.39$\pm$0.12, 56  & $-$0.63$\pm$0.09, 54 &   0.11$\pm$0.15, 53 &   0.37$\pm$0.13, 53 &               \\
$log{\tau_{\rm c}}$  & $-$0.08$\pm$0.15, 56 & $-$0.01$\pm$0.14, 56  &    0.26$\pm$0.15, 54 &   0.04$\pm$0.16, 53 &   0.03$\pm$0.15, 53 &                \\
$log{\tau_{\rm hm}}$ & $-$0.03$\pm$0.15, 56 &    0.26$\pm$0.13, 56  & $-$0.14$\pm$0.15, 54 &   0.20$\pm$0.16, 53 &   0.17$\pm$0.16, 53 &                 \\
R$_{\rm GC}$       & $-$0.29$\pm$0.14, 56 &    0.02$\pm$0.14, 56  & $-$0.05$\pm$0.13, 54 &   0.02$\pm$0.16, 53 &$-$0.24$\pm$0.13, 55 &                 \\
age (MF09)         & $-$0.25$\pm$0.13, 55 & $-$0.01$\pm$0.14, 55  &    0.11$\pm$0.15, 53 &$-$0.36$\pm$0.11, 52 &$-$0.08$\pm$0.12, 52 &                \\
age (D10)          & $-$0.30$\pm$0.12, 55 &    0.16$\pm$0.15, 55  &    0.08$\pm$0.13, 53 &$-$0.24$\pm$0.13, 52 &$-$0.10$\pm$0.13, 52 &                \\
age (V13)          & $-$0.53$\pm$0.10, 51 &    0.02$\pm$0.16, 51  &    0.06$\pm$0.14, 49 &$-$0.49$\pm$0.11, 49 &$-$0.23$\pm$0.14, 49 &                \\
 {\rm [Fe/H]}      &    0.67$\pm$0.08, 56 & $-$0.08$\pm$0.15, 56  & $-$0.08$\pm$0.15, 54 &   0.45$\pm$0.13, 53 &   0.47$\pm$0.12, 53 &                \\
M$_{\rm V}$        & $-$0.43$\pm$0.12, 56 & $-$0.63$\pm$0.09, 56  &    0.72$\pm$0.07, 54 &$-$0.38$\pm$0.13, 53 &$-$0.59$\pm$0.10, 53 &                 \\   
log{$\mathcal{M/M_{\odot}}$} &    0.58$\pm$0.14, 43 &    0.64$\pm$0.09, 43  & $-$0.81$\pm$0.05, 43 &   0.41$\pm$0.13, 42 &   0.72$\pm$0.09, 42 &             \\
$f^{\rm C}_{\rm bin}$   &    0.17$\pm$0.17, 34 & $-$0.36$\pm$0.14, 34  &    0.50$\pm$0.17, 33 &$-$0.02$\pm$0.20, 33 &$-$0.08$\pm$0.19, 33 &           \\
$f^{\rm C-HM}_{\rm bin}$& $-$0.11$\pm$0.16, 46 & $-$0.43$\pm$0.10, 46  &    0.58$\pm$0.11, 45 &$-$0.08$\pm$0.17, 45 &$-$0.32$\pm$0.15, 45 &           \\
$f^{\rm oHM}_{\rm bin}$& $-$0.19$\pm$0.16, 41 & $-$0.30$\pm$0.14, 41  &    0.65$\pm$0.12, 40 &   0.27$\pm$0.16, 39 &$-$0.32$\pm$0.15, 39 &            \\
$S_{\rm RR~Lyrae}$      & $-$0.25$\pm$0.13, 56 & $-$0.15$\pm$0.14, 56  & $-$0.08$\pm$0.14, 54 &   0.17$\pm$0.14, 55 &$-$0.13$\pm$0.13, 53 &            \\
$E(B-V)$ &    0.37$\pm$0.13, 56 &   0.19$\pm$0.16, 55   &    0.11$\pm$0.14, 54 &   0.07$\pm$0.15, 53 &   0.29$\pm$0.29, 53 &            \\
$N_{\rm 1}/N_{\rm TOT}$  & $-$0.36$\pm$0.13, 53 & $-$0.49$\pm$0.11, 52  &       1.00, 54       &$-$0.25$\pm$0.14, 53 &$-$0.59$\pm$0.10, 53 &                   \\
\hline
\hline
\end{tabular}\\
}
\label{tab:correlazioni}
\end{table*}
%%%%%%%%%%%%%%%%%%%%%%%%%%%%%%%%%%%%%%%%%%%%%%%%%%%%%%%%%%%%%%%%%%%%%%%%%%      

\appendix 
\section{The construction of the chromosome map of  $\omega$\,Centauri}
\label{sub:wCen}
 $\omega$ Centauri shows the most-complex chromosome map. The distribution of the stars that we have colored black in Figure~\ref{fig:maps4} resembles that of some GCs with single SGB like NGC\,6723 or NGC\,2808. In contrast, red-RGB stars exhibit an unique pattern,  with three main streams of red-RGB stars. The most populous RGB starts from  ($\Delta_{\rm F275W,F814W};\Delta_{\rm F275W,F336W,F438W}$)$\sim$($-0.2;0.35$) and extends towards extreme values of $\Delta_{\rm F275W,F814W} \sim 1.5$. A second stream ranges from ($\Delta_{\rm F275W,F814W};\Delta_{\rm F275W,F336W,F438W}$)$\sim$($0.0;0.1$) to ($1.5;-0.4$) and possibly includes a few stars with even larger $\Delta_{\rm F275W,F814W}$. A third stream has intermediate $\Delta_{\rm F275W,F814W}$ and $\Delta_{\rm F275W,F336W,F438W}$ values with respect to the previous two.

Each stream includes sub-stellar populations. In an attempt to estimate how many groups of stars are statistically significant in $\omega$\,Centauri, we used the Mcluster CRAN package in the public domain R statistical software system.  This package performs a maximum likelihood fits to different number of stellar groups by using several different assumptions about shape and size of the different populations in the chromosome map, and evaluate the number of groups by the Bayesian Information Criterion (BIC) penalized likelihood measure for model complexity (see McLachlan \& Peel 2000 for details). 
 For each shape and size that we adopted for the populations, we assumed a number, N, of stellar populations from 1 to 20 and estimated a BIC for each combination. The best BIC value corresponds to N=16.

\begin{centering}
\begin{figure*}
 \includegraphics[width=8.8cm]{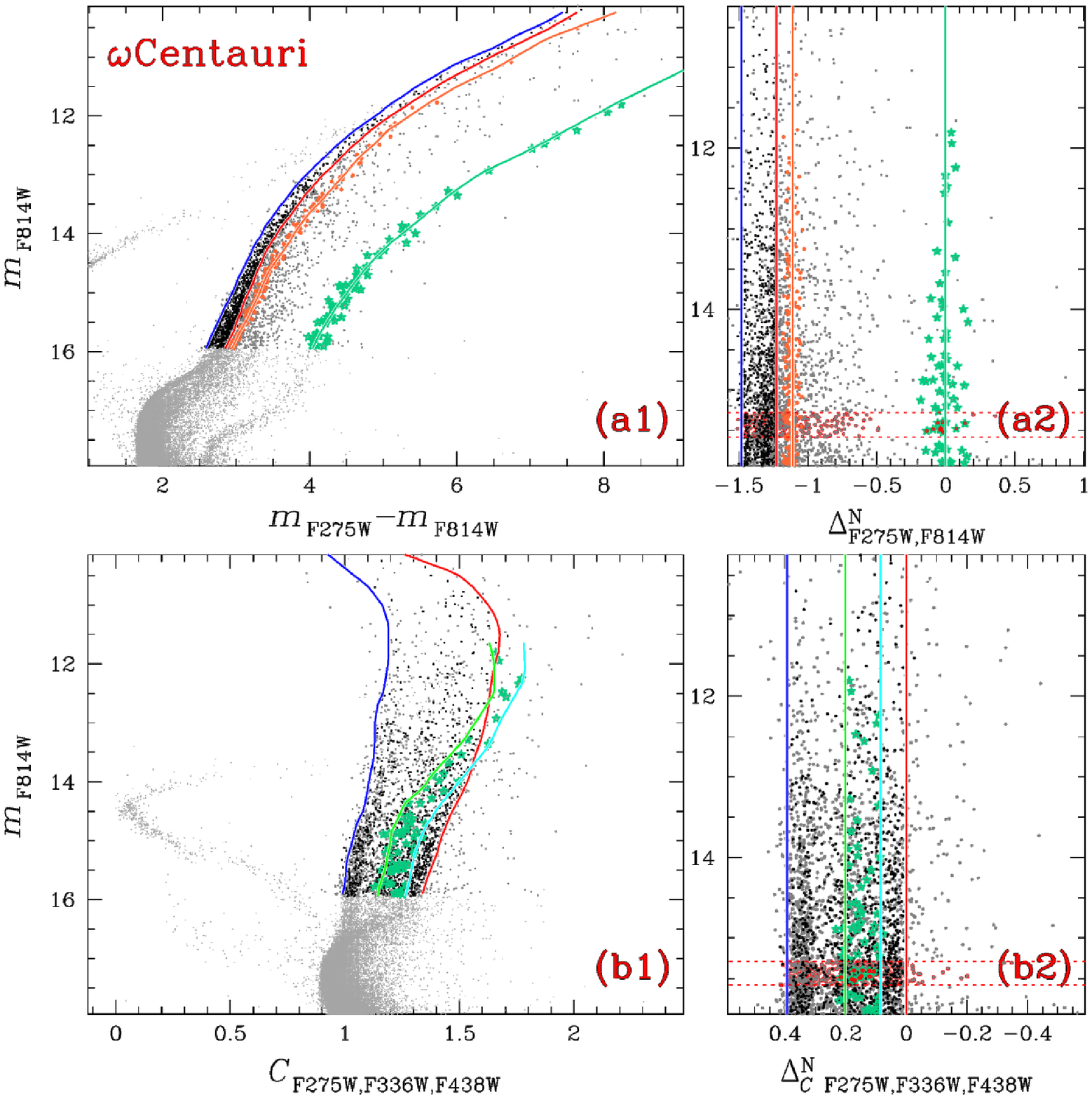}
 \includegraphics[width=8.8cm]{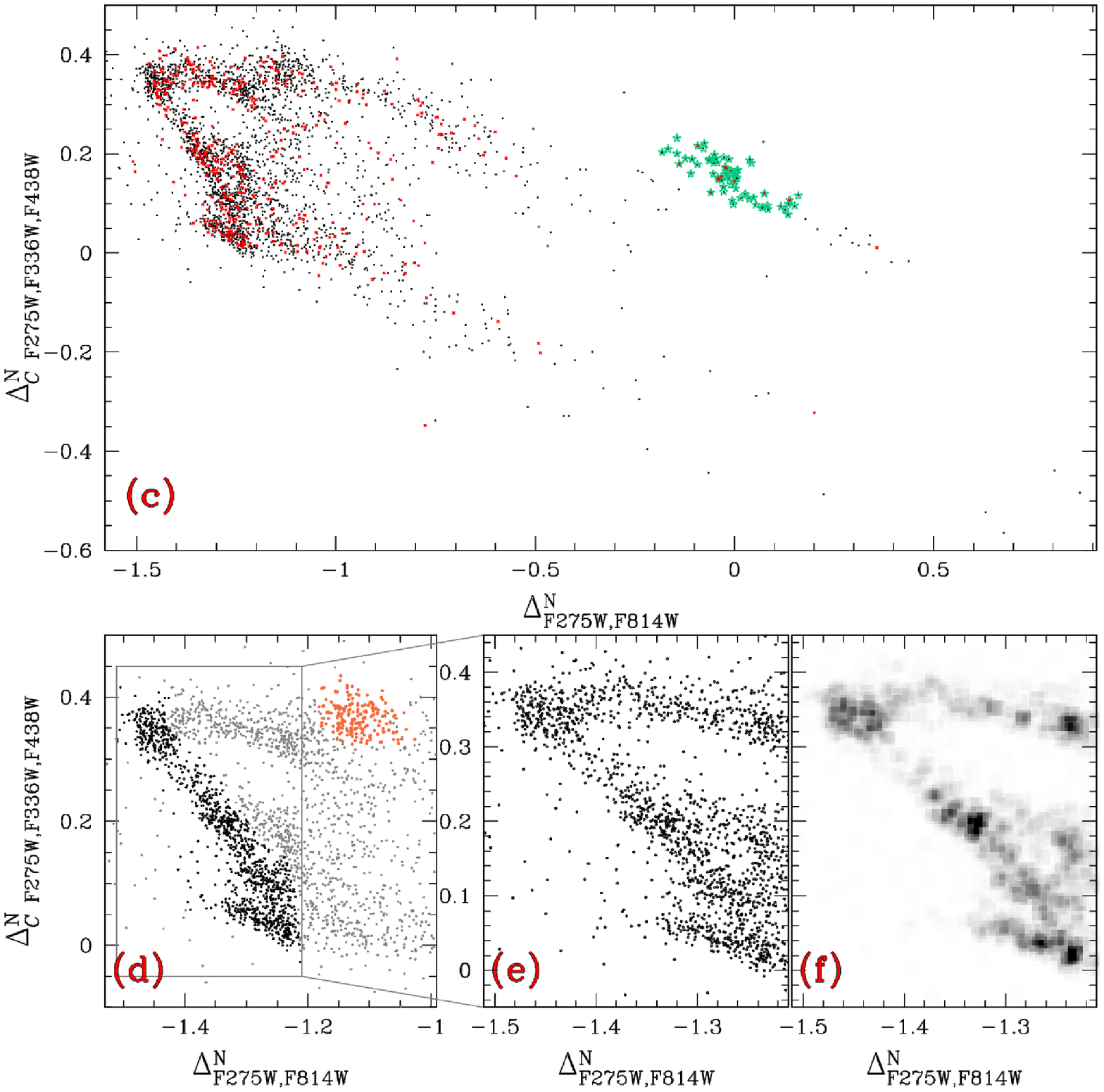}
%/home/milone/MW/GCs/NGC5139/WFC3/fig.macro fig0 fig
 \caption{This figure illustrates the procedure used to derive the chromosome map of NGC\,5139 ($\omega$\,Centauri). The $m_{\rm F814W}$ vs.\,$m_{\rm F275W}-m_{\rm F814W}$ CMD and the $m_{\rm F814W}$ vs.\,$C_{\rm F275W,F336W,F438W}$ pseudo-CMD are plotted in panels a1 and b1, respectively. Dark-gray and colored points to mark the sample of analyzed RGB stars. 
 The red and the blue lines overimposed on the diagrams of both panels a1 and d1 correspond to the red and the blue envelopes of RGB of  stars in the sample\,1. The orange and the green lines shown in the panel a1 are the fiducial lines of the sample 2 and 3 of stars.
 In the panel b1 we have used green and cyan colors to mark the red and blue edges of the envelope of RGB of sample-3 stars.
 Panels a2 and b2 show the verticalized $m_{\rm F814W}$ vs.\,$\Delta_{\rm F275W,F814W}$ and $m_{\rm F814W}$ vs.\,$\Delta_{C \rm F275W,F336W,F438W}$ diagrams for RGB stars. 
 The $\Delta_{C \rm F275W,F336W,F438W}$ vs.\,$\Delta_{\rm F275W,F814W}$ chromosome map of RGB stars in $\omega$\,Centauri is shown in panel c, where red dots represent RGB stars with $15.28<m_{\rm F814W}<15.58$ between the two horizontal dotted lines of panels a2 and b2.
 Panels d and e are zoomed-in view of the chromosome map shown in panel c, while panel f shows the $\Delta_{\rm F275W,F336W,F438W}$ vs.\,$\Delta_{\rm F275W,F814W}$ Hess diagram of the stars plotted in panel e.
 The aqua starred symbols plotted in panels a1, a2, b1, b2, and c mark the sample-3 stars, while sample-1 stars are represented with black dots in panels a1, a2, b1, b2, and d. The orange dots shown in panels a1 a2 and d indicate sample-2 stars. 
}
 \label{fig:mapWcen}
\end{figure*}
\end{centering}

When compared with the other  GCs investigated in this paper, $\omega$\,Centauri exhibits by far the most-complex CMD  and its RGB spans a very wide range of $m_{\rm F275W}-m_{\rm F814W}$ color as shown in panel a1 of Figure~\ref{fig:mapWcen}. Due to the complex structure of its RGB, in order to derive the chromosome map of $\omega$\,Centauri we have adopted an iterative procedure that is based on the method of Section~\ref{sec:mappe}, and which is illustrated in Figure~\ref{fig:mapWcen}. 

As a first step, we have derived a raw chromosome map by using the same procedure described in Section~\ref{sec:mappe}. Then, we have identified three groups of stars that have been used to derive the fiducial lines shown in $m_{\rm F814W}$ vs.\,$m_{\rm F275W}-m_{\rm F814W}$ CMD and the $m_{\rm F814W}$ vs.\,$C_{\rm F275W,F336W,F438W}$ pseudo-CMD plotted in panels a1 and b1 of Figure~\ref{fig:mapWcen}. The selected groups of stars are shown in panels c and d of Figure~\ref{fig:mapWcen} where black and orange dots and aqua starred symbols overimposed on the final chromosome map of $\omega$\,Centauri represent stars of the samples 1, 2, and 3, respectively.

 These three groups of stars have been determined iteratively by using the following criteria. The chromosome map of stars in sample 1 resembles those observed in several GCs in which the stars  are distributed along a single sequence and define distinct bumps. Sample 2 includes the bump of stars around ($\Delta_{\rm F275W,F814W}$;$\Delta_{\rm F275W,F336W,F438W}$)$\sim$($-$1.10;0.35), while sample 3 includes most of the stars of the reddest and the most metal-rich RGB of $\omega$\,Centauri that has been often indicated as population a (e.g., Bedin et al.\,2004). Noticeably, we have excluded from sample 3 the stars in the poorly populated bump with ($\Delta_{\rm F275W,F814W}$;$\Delta_{\rm F275W,F336W,F438W}$)$\sim$(0.3;0.0).

In order to derive $\Delta_{\rm F275W,F814W}$ for RGB stars in $\omega$\,Centauri, we have used the following procedure that is illustrated in panels a1 and a2 of Figure~\ref{fig:mapWcen}. 
 We have first divided the RGB stars in three groups. Group I includes all the RGB stars with bluer $m_{\rm F275W}-m_{\rm F814W}$ colors than the red fiducial line at the corresponding F814W magnitude. Group II includes the RGB stars between the red and the orange line, while the remaining RGB stars belong to group III  
 The red and the blue fiducial lines shown in panel a1 are the redder and the bluer envelopes of the RGB formed by sample-1 stars and have been derived as in Section~\ref{subsec:W} by using sample-1 stars only. The orange and the aqua lines shown in panel a1 are fiducial lines of the RGB made by sample 2 and sample 3 stars. 
We have  derived the quantities $\Delta_{\rm F275W,F814W}^{\rm N\,I}$, $\Delta_{\rm F275W,F814W}^{\rm N\,II}$, and $\Delta_{\rm F275W,F814W}^{\rm N\,III}$ for stars in the three groups, by using the following equations that are similar to Equation~\ref{eq:1}: 
\begin{equation}\label{eq:3}
\Delta_{\rm F275W,F814W}^{\rm N\,I (II, III)}= W_{\rm F275W,F814W}^{\rm I, (II, III)} [(X-X_{\rm fiducial~A})/(X_{\rm fiducial~B}-X_{\rm fiducial~A})]. 
\end{equation}

For group-I stars, we have assumed the blue and the red fiducial shown in panel a1 of Figure~\ref{fig:mapWcen} as the fiducial A and fiducial B, respectively. For group-II stars, the red fiducial corresponds to fiducial A and the orange fiducial corresponds to fiducial B, while for group-III stars we used the orange and the green fiducials as fiducial A and B, respectively.  The constant $W_{\rm F275W,F814W}^{\rm I}$ has been derived for group-I stars as in Section~\ref{subsec:W}, while $W_{\rm F275W,F814W}^{\rm II}$ has been derived as the $m_{\rm F275W}-m_{\rm F814W}$ color difference between the orange and the red fiducial line  shown in panel a1, of Fig~\ref{fig:mapWcen} calculated 2.0 F814W mag above the MS turn off. The constant $W_{\rm F275W,F814W}^{\rm III}$ has been derived similarly for group III star, but by using green and orange fiducials. 
 
We assumed:\\
 $\Delta_{\rm F275W,F814W}=\Delta_{\rm F275W,F814W}^{\rm N\,I}$ for group I stars;\\
 $\Delta_{\rm F275W,F814W}=W_{\rm F275W,F814W}^{\rm I}+\Delta_{\rm F275W,F814W}^{\rm N\,II}$ for group-II stars; and\\
 $\Delta_{\rm F275W,F814W}=W_{\rm F275W,F814W}^{\rm I}+W_{\rm F275W,F814W}^{\rm II}+\Delta_{\rm F275W,F814W}^{\rm N\,III}$ for group-III stars. 
 The verticalized $m_{\rm F814W}$ vs.\,$\Delta_{\rm F275W,F814W}$ diagram of the analyzed RGB stars in $\omega$\,Centauri is plotted in the panel a2 of Figure~\ref{fig:mapWcen} where the vertical colored lines corresponds to the fiducial lines shown in panel a1.

In order to derive $\Delta_{C~ \rm F275W,F336W,F438W}$ for RGB stars of  $\omega$\,Centauri we adopted the method illustrated in panels b1 and b2 of  Fig~\ref{fig:mapWcen}, where the red and blue lines are the boundaries of the RGB for stars in  sample 1, while the green and the cyan lines are the boundaries for stars in the sample 3. These lines have been derived as described in Section~\ref{subsec:W}.

 We proceeded by defining two additional groups of stars. Group IV includes all the RGB stars that are associated with the most-metal-rich population of $\omega$\,Centauri and that have $\Delta_{F275W,F814W}>-0.2$ and $\Delta_{C \rm F275W,F336W,F438W}>-0.1$ in panel c of Figure~\ref{fig:mapWcen}. Group V includes all the remaining RGB stars.

 We derived $\Delta^{\rm N, IV}_{C \rm F275W,F336W,F438W}$ for group-IV stars by means of Eq.~\ref{eq:2} and by assuming the green and cyan lines plotted in panel b1 of Figure~\ref{fig:mapWcen} as fiducials A and B, respectively. Similarly, we have calculated $\Delta^{\rm N, V}_{C \rm F275W,F336W,F438W}$ by using Eq.~\ref{eq:2} and assuming that the blue and red lines in the panel b1 of Figure~\ref{fig:mapWcen} correspond to the fiducials A and B, respectively.

We assumed:\\
$\Delta_{C \rm F275W,F336W,F438W}=\Delta^{\rm N, IV}_{C \rm F275W,F336W,F438W}$ for group-IV stars and;\\
$\Delta_{C \rm F275W,F336W,F438W}=W_{C, \rm F275W,F336W,F438W}^{\rm IV-V}+\Delta^{\rm N, V}_{C \rm F275W,F336W,F438W}$ for group-V stars, where $W_{C, \rm F275W,F336W,F438W}^{\rm IV-V}$ is the $C_{\rm F275W,F336W,F438W}$ pseudo-color difference between the blue and the cyan fiducial line calculated 2.0 F814W magnitudes above the MS turn off.
  
The chromosome map has been derived iteratively and four iterations were required to reach convergence. After each iteration, we improved the selection of stars in the samples 1, 2, and 3, derived improved fiducial lines and better estimates of $\Delta_{\rm F275W,F814W}$ and $\Delta_{C \rm F275W,F336W,F438W}$.   

%%%%%%%%%%%%%%%%%%%%%%%%%%%%%%%%%%%%% FIG 2 %%%%%%%%%%%%%%%%%%%%%%%%%%%%%%%%%%%
%%%%%%%%%%%%%%%%%%%%%%%%%%%%%%%%%%%%%%%%%%%%%%%%%%%%%%%%%%%%%%%%%%%%%%%%%%%%%%%

The chromosome map of $\omega$\,Centauri is plotted in the panel c of Figure~\ref{fig:mapWcen} and it reveals a very complex stellar distribution, with the presence of distinct bumps of stars and stellar streams. 
 To verify that the observed structure do not include artifacts introduced by the adopted fiducial lines, we marked in red in panel c all the stars in a small magnitude interval with $15.38<m_{\rm F814W}<15.58$ that are placed between the horizontal dotted lines of panels a2 and b2.  The fact that the selected stars distribute along the entire map demonstrates that the observed stellar bumps and tails are real.
 Panels d and e of Figure~\ref{fig:mapWcen} are a zoomed-in view of the chromosome map around the region with low values of $\Delta_{\rm F275W,F814W}$, while panel f shows the Hess diagram of the same stars plotted in panel e.  These figures reveal that the sample-1 of stars in $\omega$ Centauri define a continuous sequence characterized by the presence of distinct stellar bumps, in close analogy with what we observe in NGC\,6723. In addition, $\omega$ Centauri hosts stellar populations, including bumps and streams, with values of $\Delta_{\rm F275W,F814W}$ larger than those of sample-1 stars with the same  $\Delta_{\rm F275W,F336W,F438W}$. 

 \section*{acknowledgments}
\small
 Support for Hubble Space Telescope proposal GO-13297 was provided by NASA through grants from STScI, which is operated by AURA, Inc.,  under NASA contract NAS 5-26555.
We thank the anonymous referee for her/his suggestions that have improved the quality of the paper.
 APM and AFM acknowledge support by the Australian Research Council through Discovery Early Career Researcher Awards DE150101816 and and DE160100851. GP, AR, FD, SC acknowledge financial support by PRIN-INAF2014 (PI: Cassisi).
\bibliographystyle{aa}

\bibliographystyle{aa}

\end{document}